\documentclass[useAMS,usenatbib,twocolumn]{mn2e}
\usepackage{epsfig}
\usepackage{amsmath}
\newcommand{\be}{\begin{equation}}
\newcommand{\beq}{\begin{equation}}
\newcommand{\ba}{\begin{eqnarray}}
\newcommand{\ee}{\end{equation}}
\newcommand{\eeq}{\end{equation}}
\newcommand{\ea}{\end{eqnarray}}

\def\lsim{~\rlap{$<$}{\lower 1.0ex\hbox{$\sim$}}}

\def\gsim{~\rlap{$>$}{\lower 1.0ex\hbox{$\sim$}}}

\title[Large Gpc volume reionization simulations]{A hybrid multi resolution scheme to {efficiently} model the structure of reionization on the largest scales}
\author[Han-Seek Kim et al.]
       {Han-Seek~Kim$^{1}$\thanks{hansikk@unimelb.edu.au}, J. Stuart B.~Wyithe$^{1,2}$, Jaehong~Park$^{1}$, Gregory B. Poole$^{1}$, 
              \vspace{0.3cm} \\ 
      \LARGE\rm{C. G.~Lacey$^{3}$, and C. M.~Baugh$^{3}$} \\
       $^1$School of Physics, The University of Melbourne, Parkville, VIC 3010, Australia\\
      $^2$ARC Centre of Excellence for All-sky Astrophysics (CAASTRO)\\
      $^3$Institute for Computational Cosmology, Department of Physics, University of Durham, South Road, Durham DH1 3LE, UK}
  	
\date{}

\pagerange{\pageref{firstpage}--\pageref{lastpage}}
\pubyear{2015} 

\begin{document}

\maketitle

\label{firstpage}

\begin{abstract}
Redshifted 21cm measurements of the structure of ionised regions that grow during reionization promise to provide
a new probe of early galaxy and structure formation. One of the challenges of modelling reionization is to account both for the sub-halo scale physics of galaxy formation and the regions of ionization on scales that are many orders of magnitude larger. To bridge this gap we first calculate the statistical relationship between ionizing luminosity and Mpc-scale overdensity using detailed models of galaxy
formation computed using relatively small volume - ($\sim$100Mpc/$h$)$^{3}$, high resolution dark matter simulations. We then use a Monte-Carlo technique to apply this relationship to reionization of the intergalactic medium within large volume dark matter simulations - ($>$1Gpc/$h$)$^{3}$. The
resulting simulations can be used to address the contribution of very large scale clustering of galaxies to the structure of reionization, and show that volumes larger than 500Mpc/$h$ are required to probe the largest reionization features mid-way through reionization. As an example application of our technique, we demonstrate that the predicted 21cm power spectrum amplitude and gradient could be used to determine the importance of supernovae feedback for early galaxy formation.
\end{abstract}


\section{Introduction}
A new generation of radio telescopes including LOFAR\footnote{http://lofar.org} (LOw Frequency Array), MWA\footnote{http://haystack.mit.edu/arrays/MWA} (Murchison Widefield Array), and PAPER\footnote{http://eor.berkeley.edu} (Precision Array for Probing the Epoch of Reionization) hope to observe the evolution of neutral hydrogen during the reionization of the Universe. The resulting measurements of the timing and structure of reionization promise to probe the properties of the first galaxies \citep{BL01,pen2009,mesinger2010,Ahn2012}.

Theoretical modelling suggests that on large scales overdense regions are reionized first  due to galaxy bias \citep{ciardi2003,furl2004a,furl2004b,WM07,iliev2007,ZLMDHZF07,mcquinn2007,trac2008}. The size and evolution of HII regions is therefore sensitive to the process of galaxy formation because the distribution of ionizing photons relative to the density field depends on the typical halo mass of star forming galaxies. For example, there has been a range of studies which show that reionization can be self-regulating \citep{dijkstra2004, iliev2007, mesinger2008, Ahn2012} because low mass galaxies are suppressed in a heated IGM. 
Supernova feedback also plays a significant role in the history and structure of reionization by suppressing star formation in lower mass halos \citep{WL13, Kim2012a}. 

Simulations of large volumes of the IGM during reionization are important for interpreting upcoming observational programs with the MWA and LOFAR because of their large field of view, which correspond to several Gpc at $z>6$. In addition, large volume simulations are essential for  testing of convergence of reionization properties \citep{iliev13}. However, until very recently, the largest simulations that include physical modelling of galaxy formation had a box size of $\sim$100Mpc  \citep{Kim2012a,Norman2013,Gnedin2014,Ilustris2014}. Larger volumes have generally employed fully semi-numerical schemes or radiative transfer based on simple source models for the relationship between the ionizing luminosity and host dark matter halo mass \citep{S+10, mesinger2010,iliev13}. Recently, \cite{Battaglia2013} suggested a method for calculating the evolution of the 3-dimensional ionization field  in $>$ (Gpc/$h$)$^{3}$ volumes using the correlation between the ionization field and dark matter overdensity field at different redshifts from high resolution radiation-hydrodynamic
simulations. This method accurately reproduces the ionization structure on the scales tested but does not show an increase in large scale power when the box size is increased, as has been shown in the direct simulations of \cite{iliev13}.

In this paper we introduce a new method to perform very large volume ($>$ Gpc/$h$ box size) reionization simulations, whilst modelling the galaxy formation physics using smaller volumes (100Mpc/$h$ box size). Our model is based on the GALFORM galaxy formation model \cite[]{Bower2006, Lagos2012}. We employ GALFORM within the Millennium-II simulation \cite[]{MII2009}, and combine it with a semi-numerical scheme to calculate the structure of reionization as described in \cite{Kim2012a}. We begin in \S\ref{modell} by briefly describing the implementation of GALFORM, and our method for simulating reionization. Then, in \S\ref{Gmodel} we describe our method for translating the galaxy formation physics to large volume reionization simulations. We discuss some implications in \S\ref{Imp}, and finish with our Summary in \S\ref{Summary}.

%\section{The models} \label{modell}
\section{A semi-numerical model for reionization} \label{modell}
In this section we briefly introduce reionization modelling based on the method described in \cite{Kim2012b}. We combine the semi-analytic galaxy formation model GALFORM (\S\ref{GFM}) with {an improved} semi-numerical scheme (\S\ref{scheme}) to generate an ionization field. %In \S\ref{scheme} we also describe an improved method to find the ionized structure, and 
In \S\ref{PS} we present the resulting redshifted 21-cm power spectrum.

\subsection{The GALFORM galaxy formation model}\label{GFM}

The GALFORM semi-analytic galaxy formation model successfully explains  a large range of observed properties of galaxies at low redshifts \citep[][]{Kim2011, Kim2012, Lagos2012, Kim2012a, Kim2015HI}. GALFORM includes a range of processes that are thought to be important for galaxy formation \citep[see][for more details]{Cole2000, Baugh2006, Bower2006,Lagos2012}. 
In this paper, we implement GALFORM in halo merger trees extracted from the Millennium-II cosmological N-body simulation \citep[][]{MII2009}; see \cite{Jiang2014} for a description of the construction of merger trees. The Millennium-II simulation has a cosmology with fractional mass and dark energy densities values of $\Omega_{\rm m}=0.25$, $\Omega_{\rm b}=0.045$ and $\Omega_{\Lambda}$=0.75, a dimensionless Hubble constant of $h$=0.73, and a power spectrum normalisation of $\sigma_{8}$=0.9 (Millennium cosmology for table~\ref{Nbodies}). The resolution of the simulation is fixed at a halo mass of $\sim$10$^{8}$${\rm M_{\odot}}$/h in the simulation box of side length $L$=100Mpc/$h$. 
Note that we use the \cite{Lagos2012} implementation of GALFORM for this paper. 
%%%%%%%%%%%%%%

\subsection{Semi-Numerical scheme}\label{scheme}

We use semi-numerical modelling \citep[e.g.][]{MF07,GW08,ZLMDHZF07} which is an approximate but efficient method for
simulating the reionization process. 
%{\bf Since we have information about the positions of dark matter particles and galaxies, f}ollowing \cite{Kim2012b}  
{Because our modelling is based on the Millennium-II simulation, which has positional information for dark matter haloes and galaxies,}
we begin by gridding the ionizing luminosities of galaxies from the GALFORM model into small volumes (or cells). We assume the number of photons produced by galaxies in the cell that enter the IGM and participate in reionization to be
\begin{equation}\label{nphotons}
N_{\rm \gamma, cell}={\it f}_{\rm{esc}}\int^{t_{z}}_{0}\dot{N}_{\rm Lyc,cell}(t)~{\rm{d}}t,
\end{equation}
where $f_{\rm esc}$ is the escape fraction of ionizing photons produced by stars in a galaxy {and $t_{z}$ is the age of the Universe at redshift $z$}. The total Lyman continuum luminosity of the $N_{\rm cell}$ galaxies within the cell, expressed as the rate of emission of ionizing photons { (i.e.  units of photons/s)}, computed from GALFORM is
\begin{equation}
\dot{N}_{\rm Lyc,cell}(t) = \sum_{i=1}^{N_{\rm cell}} \dot{N}_{\rm Lyc, \it i}(t),
\end{equation}
where
\begin{equation}
\dot{N}_{\rm Lyc, \it i}(t)=\int^{\infty}_{\nu_{\rm thresh}}{L_{\nu,i}(t) \over h\nu} {\rm d\nu},
\end{equation}
L$_{\nu,i}$ is the spectral energy distribution of galaxy $i$, and $\nu_{\rm thresh}$ is the Lyman-limit frequency, $h\nu_{\rm thresh}$ = 13.6 eV.

We then calculate the {ionized hydrogen fraction} within each cell according to  
\begin{equation}
\label{Qvalue}
Q_{\rm cell}=\left[{N_{\rm \gamma, cell} \over (1+F_{\rm c})N_{\rm H, cell}}\right],
\end{equation}
where $F_{\rm c}$ denotes the mean number of recombinations per hydrogen atom up to reionization and $N_{\rm H, cell}$ is the number of hydrogen atoms within a cell. {We choose the values $f_{\rm esc}$ and $F_{c}$ to get a similar evolution of {mean global mass averaged ionized hydrogen} fraction to the one shown in \cite{lidz2008} \cite[see detailed values in][] {Kim2012a}.} {We note that our assumption is that values of $F_{\rm c}$ and $f_{\rm esc}$ do not depend on the galaxy mass or redshift. In reality the escape fraction may be mass and redshift dependent, and the mean number of recombinations per hydrogen atom may be dependent on the overdensity of intergalactic medium \citep{Inoue2006, gnedin2007a, WM07, WC09, kuhlen2012, yajima2011, Kim2013}}. The latter quantity is calculated as 
\begin{equation}
\label{nHI}
N_{\rm H, cell}={n_{\rm H}(\delta_{\rm dm,cell}+1)V_{\rm cell}},
\end{equation}
where we assume that the overdensity of {hydrogen atoms} follows the dark matter (computed based on the Millennium-II simulation density field, 1+$\delta_{\rm dm,cell}=\rho_{\rm dm,cell}/{\bar \rho_{\rm dm}}$), $n_{\rm H}$ is the mean comoving number density of hydrogen atoms, and $V_{\rm cell}$ is the comoving volume of the cell. Self-reionization of a cell occurs when $Q_{\rm cell}>1$. We divide the Millennium-II simulation box into either 256$^{3}$ or 50$^{3}$ cells, yielding cell side lengths of 0.3906Mpc/$h$ or 2Mpc/$h$, and comoving volumes of 0.0596Mpc$^{3}$/h$^{3}$ or 8Mpc$^{3}$/h$^{3}$ respectively.

Since $Q_{\rm cell}$ can take a value greater than 1, radiation from a cell with $Q_{\rm cell}>1$ can ionize a neighbouring cell with $Q_{\rm cell}<1$. In order to find the extent of ionized regions we therefore filter the %$Q_{\rm cell}$ 
{ ${N_{\rm \gamma, cell}}$ and $N_{\rm H, cell}$} fields using a sequence of real-space top hat filters of radius $R$ (from the cell size to box size), producing one smoothed ionization field {$Q_{R}$ per radius calculated by
\begin{equation}
\label{QvalueR}
Q_{R}=\left[{N_{\rm \gamma, R} \over (1+F_{\rm c})N_{\rm H, R}}\right],
\end{equation}
where $N_{\rm \gamma, R}$ ($N_{\rm H, R}$) is the sum of the number of photons (sum of number of hydrogen atoms) in a sphere of radius R.}  
At each point in the
simulation box, we find the largest $R$ for which the filtered ionization field is greater than unity (i.e. ionized with $Q_{R} >$ 1). All cells within radius $R$ around this point are considered ionized. We also include partial ionization for cells (from {\bf Eq.~\ref{Qvalue}}). 

\begin{figure}
\begin{center}
\includegraphics[width=8cm]{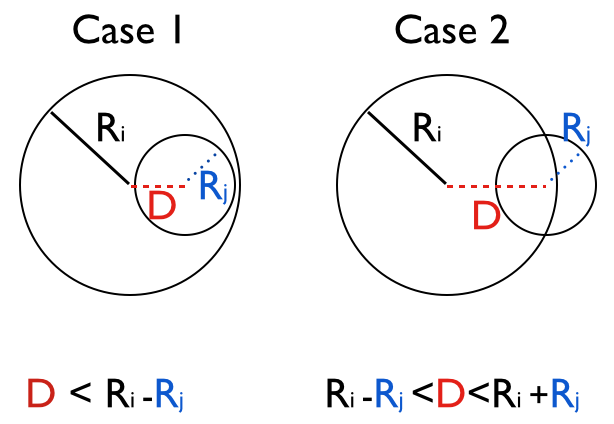}
\end{center}
\vspace{-5mm}
\caption{The different two cases of overlap between two HII bubbles. R$_{i}$ and R$_{j}$ are the radii of two individual bubbles and D is the distance between the centres of the bubbles ${i}$ and ${j}$.} \label{Cases}
\end{figure}

\begin{table}
\caption{
The values of the expected mean global mass averaged ionized hydrogen fractions, $\left<x_{\rm i}\right>$, from the semi-analytic {model} for different redshifts (selected for comparison with the work by \citet{lidz2008}) and values of the expected mean global mass averaged neutral hydrogen fractions, $\left<x_{\rm HI}\right>$. Results of the values of mean mass averaged neutral hydrogen fraction, $\left<x_{\rm HI,Semi}\right>$, from the semi-numerical scheme for different redshifts. This case assumed the default model with Millennium-II and the \citet{Lagos2012} GALFORM model.}
\label{Fractions}
\begin{tabular}{ccccccc}
\hline
\hline
Redshift (z) & 9.278&8.550&7.883&7.272&6.712&6.197\\
\hline
$\left<x_{\rm i}\right>$ & 0.056&0.16&0.36&0.55&0.75&0.95\\
$\left<x_{\rm HI}\right>$ & 0.944&0.84&0.64&0.45&0.25&0.05\\
\hline
$\left<x_{\rm HI,Semi}\right>$  & 0.98& 0.85&0.67& 0.47& 0.25&0.059\\
\hline
\end{tabular}
\end{table}
Our method treats each cell as a source. To find HII regions which properly conserve photons from sources when the HII regions overlap, we take the following steps \citep{ZLMDHZF07, thomas2009}. We use real space spherical filtering, and so have information regarding which HII bubbles overlap (this is not possible in Fourier space). When filtering we start with the smallest radius corresponding to the size of cell and increase to the size of simulation box (increasing the filtering radius in linear intervals). To properly include overlap between HII regions in the semi-numerical scheme\footnote{Note that \citet{Kim2012b} used real space top hat filters of radius from the box size to the cell size. The filtering from large radius to small radius resulted in double counted photons in the overlap regions of neighbouring bubbles, and so the model did not satisfy photon conservation. Our calculations in this paper improve photon conservation relative to the method in \citet{Kim2012b}.}, we consider two cases (shown schematically in Fig.~\ref{Cases}).
We refer to the cell at the centre of region $i$ with radius $R_{\rm i}$ as the main cell. 
  
Case 1 : Cells $i$ $\&$ $j$ separated by distance D have bubble radii such that bubble $j$ is enclosed within bubble $i$ (R$_{i} > $R$_{j}$). In this case we add all photons when calculating $Q_{R_{\rm i}}$.

\begin{figure}
\begin{center}
\includegraphics[width=8cm]{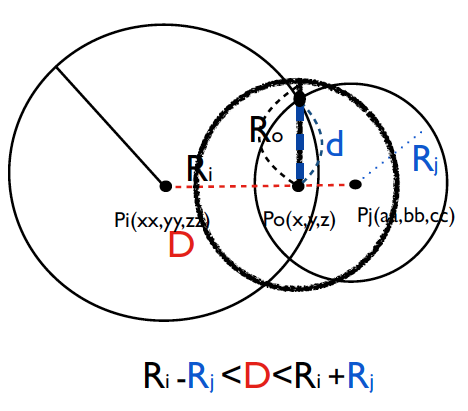}
\end{center}
\vspace{-5mm}
\caption{The semi-numerical scheme to include overlap between two HII bubbles. R$_{i}$ and R$_{j}$ are the radii of two individual bubbles and D is the distance between the centres of the bubbles ${i}$ and ${j}$. The third HII bubble (radius R$_{o}$) is centred at a point of internal division ({\it P$_{o}$}) between the two HII bubbles in the overlap area.} \label{Solve}
\end{figure}

Case 2: The separation $D$ between two cells is smaller than the sum of their two bubble radii. This case corresponds to the partial overlap of neighbouring HII bubbles. To conserve the number of photons from cells in this case, we follow previous work which noted that photons inside the region of overlap between two HII bubbles may not increase the individual sizes of the two HII individual bubbles \citep{ZLMDHZF07, thomas2009}. Instead, these photons are likely to ionize an additional volume near the intersection between the two HII bubbles. To model this overlap, we have used a semi-numerical scheme to initially  find the two HII regions. %Based on the positions and radii of these HII bubbles we add a third HII bubble centred at a point %of internal division
%({\it P$_{o}$}) between the two HII bubbles in the overlap area. This third bubble has a radius {\it R$_{o}$}, which is related to a distance d {\bf (the radius of circle of intersection plane of two HII bubbles)}, a distance between the point of internal division ({\it P$_{0}$}) and the interaction point ({\it I$_{0}$}).
Based on the positions and radii of these HII bubbles, we add a third bubble centred at {\it P$_{0}$} and of radius {\it R$_{0}$} (see Fig.~\ref{Solve}). {\it P$_{0}$} is defined to be the
centre of the circle of intersection of the two bubbles, and we define
{\it R$_{0}$}= {\it F$_{\rm overlap}$}$\times$d, where d is the radius of this circle. {\it F$_{\rm overlap}$} is a free parameter%, corresponding to a approximate volume equivalent to that of the overlap region (Fig.~\ref{Solve}). We use {\it R$_{o}$=F$_{\rm overlap}$$\times$d} in this work. {\it F$_{\rm overlap}$} is a free parameter 
, {and} we use {\it F$_{\rm overlap}$}=1.2, which results in approximate photon conservation across the redshift range. %{Note that the value {\it F$_{\rm overlap}$} is most important for driving the highly ionized latter stage of reionization. Therefore we choose the value to match the neutral fraction at z=6.197 ($1-\left<x_{i}\right>$$\sim$0.05) for the model.} 
We ionize all cells within the third bubble. 
{To treat the case of more than 2 overlapping bubbles, we span all possible overlapping regions between all sources. We check for double counting of photons during this process by neglecting already accounted for ionizing sources.} 

{%It is worth to introduce the definitions for neutral (ionized) hydrogen fractions in this work. The global mass averaged ionized hydrogen fraction is defined by $x_{\rm i}$=$n_{\rm HII}$/ $(n_{\rm HII}+n_{\rm HI})$, where $n_{\rm HII}$ and $n_{\rm HI}$ are the number densities of HII (ionized hydrogen) and HI (neutral hydrogen), respectively. 
Based on our assumption for escape fraction and $F_{c}$, we calculate the expected mean global mass averaged ionized hydrogen fraction from the ratio between ionizing photons and hydrogen atoms. $\left<x_{\rm i}\right>$=$N_{\rm \gamma, tot} / \left[(1+F_{\rm c})N_{\rm H, tot}\right]$, where $N_{\rm \gamma, tot}$ ($N_{\rm H, tot}$) is the sum of the number of photons (sum of the number of hydrogen atoms) in the simulation. The expected mean global mass averaged neutral hydrogen fraction is then obtained from the relation $\left<x_{\rm HI}\right>$=1-$\left<x_{\rm i}\right>$. We also calculate the neutral hydrogen fraction resulting from the semi-numerical scheme by averaging over the ionization state in the simulation volume ($\left<x_{\rm HI,Semi}\right>$). If the model is working correctly, $\left<x_{\rm HI}\right>$=$\left<x_{\rm HI,Semi}\right>$, and the semi-numerical scheme perfectly conserves photons.}

\begin{figure}
\begin{center}
\includegraphics[width=8.5cm]{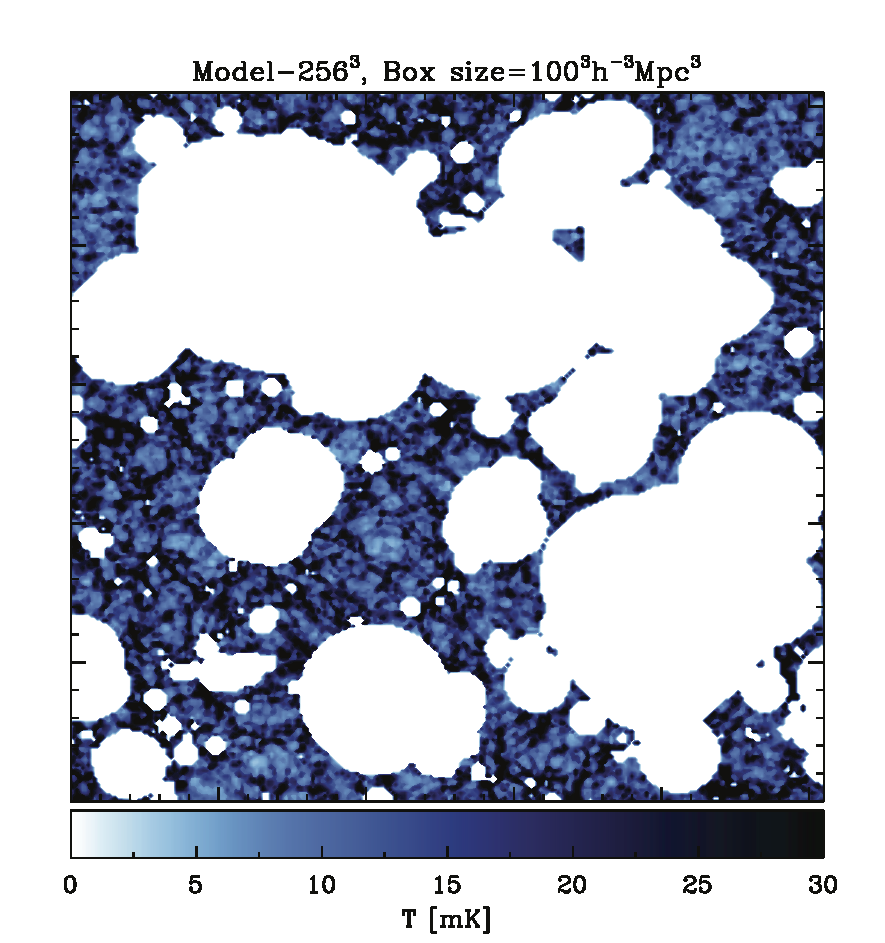}
\end{center}
\caption{The 21cm intensity map from the Model-256$^{3}$ (cell size 0.39Mpc/$h$) {at $z \sim$7.272 ($\left<x_{\rm i} \right>$$\sim$0.55)} with slice that is 0.39Mpc/$h$ deep. The colour shading shows the 21cm intensity in temperature units, as indicated by the bar.
} \label{DODED256Only}
\end{figure}

An example calculation of the ionization structure from the Millennium-II simulation and  GALFORM model \citep{Lagos2012} is shown in Fig.~\ref{DODED256Only}. To illustrate the conservation of ionizing photons in our model, Table~\ref{Fractions} shows the {mean} mass averaged neutral hydrogen fractions, $\left<x_{\rm HI,Semi}\right>$, from the semi-numerical output, together with the {expected mean global mass averaged ionized (neutral) hydrogen fraction, $\left<x_{\rm i}\right>$ ($\left<x_{\rm HI}\right>$) from the semi-analytic model for different redshifts}. {The mean mass averaged neutral hydrogen fractions using the semi-numerical scheme} agree well with the values of $\left<x_{\rm HI}\right>$, with less than 5 percent variance across the range of redshifts.

\subsection{Redshifted 21-cm intensity and power spectrum}\label{PS}
We next consider predictions for the 21cm power spectrum.
In this paper we restrict our attention to analyses that assume the spin temperature of hydrogen is coupled to the kinetic temperature of an IGM that has been heated well above the CMB temperature \citep[$z\la9$ $\&$ $T_{\rm s}\gg T_{\rm CMB}$, see][]{S+07}. This restriction is a limitation of the semi-numerical model in \cite{Kim2012a}. However we note that the method described in this paper to extend the statistics in a small simulation to larger volumes could incorporate more sophisticated models. In this regime, ignoring the contribution to the amplitude from velocity gradients and assuming the hydrogen overdensity follows the dark matter (1+$\delta_{\rm dm,cell}$), there is a proportionality between the ionized hydrogen fraction and 21~cm intensity. The 21~cm brightness temperature contrast may therefore be written as
\begin{equation}
\label{Tb}
\Delta T(z)=T_{0}(z)\left[1-Q_{\rm cell}\right]\left(1+\delta_{\rm dm,cell}\right),
\end{equation} 
where $T_{0}(z)=23.8\sqrt{(1+z)/10}$ $\mbox{mK}$.  The filtering procedure described above provides  3-dimensional maps of the ionization structure, and therefore allows us to calculate the 21-cm intensity within the simulation volume. From this we calculate the dimensionless 21-cm power spectrum 
\begin{equation}
\Delta^{2}(k)=k^3/(2\pi^2)P_{21}(k,z)/{T_{0}(z)}^{2}
\end{equation}
as a function of spatial frequency $k$, where $P_{21}(k)$ is the 3-dimensional power spectrum of 21-cm brightness temperature $\Delta T(z)$   (described by eq.~(\ref{Tb})).

The predicted power spectrum for the default model is shown as the solid curve in the top (bottom) right panel of Fig.~\ref{DOD256} at z=7.272 (7.883)\footnote{Note that we plot the power spectrum for wavenumbers less than $\sim$ 0.6 $k_{N}$ where $k_{N}$ is the Nyquist frequency of the grid to avoid the features introduce by mass assignment in a grid (cf. \cite{Cui2008}).}.
We include a statistical error on the power spectrum calculated as the uncertainty 
${\sigma(k)}= \sqrt{2 \over n_{\rm modes}}\times \Delta^{2}(k)$,
where the $n_{\rm modes}$ is the number of Fourier modes present in a spherical shell of width $\delta k$ within volume of $V$. For large scales, $k \ll 2\pi/V^{1/3}$, $n_{\rm modes}=V4\pi k^{2} \delta k/(2\pi)^{3}$, where $\delta k = 2\pi/V^{1/3}$. 

\section{Reionization in a large volume simulation.}\label{Gmodel}
{In the previous section, we introduced a semi-numerical model for reionization based on GALFORM and the Millennium-II simulation. Although simulations continue to increase in size, the method is therefore limited to volumes in which halo masses can be included down to the lowest masses thought to be responsible for reionization.}

{However, larger} volume reionization simulations are needed both to make mock observations for understanding forthcoming observations of the epoch of reionization (EoR), and also to correctly describe the amplitude of the redshifted 21cm power spectra at large scales.

\cite{iliev13} used radiative transfer to {study reionization within} a very large volume simulation. \cite{iliev13} show that the large scale power spectrum does not converge unless box sizes as large as 425Mpc/$h$ are used. Because such large volume simulations are very expensive, a method to use large volume intermediate resolution simulations from smaller volume high resolution simulations was introduced by \cite{Battaglia2013}. \cite{Battaglia2013} extract the correlation between the ionization field and dark matter overdensity field as a function of redshift using a high resolution radiation-hydrodynamic simulation. They then construct a parametric function for the bias which is used to filter a large scale density field to derive the corresponding large scale spatially varying reionization-redshift field. This method to produce large volume reionization simulations is fast. However, the results in \cite{Battaglia2013} do not show the difference between large volume and small volume calculations of the 21cm power spectrum amplitude at large scales that is seen in the simulations of \cite{iliev13}.

Therefore, we suggest a method to simulate a large volume to study reionization which has a sophisticated galaxy formation model to follow ionizing sources, is reasonably fast, and correctly calculates the amplitude of the power spectrum on large scales. 
To describe the contribution of small galaxies during the EoR, we need a dark matter simulation which can resolve sources in $\sim$10$^{8}{\rm M_{\odot}}$/h halos which are thought to dominate the production of ionizing photons \citep{iliev2007}. For this reason we have combined the GALFORM semi-analytic galaxy formation model with our semi-numerical scheme to simulate HII region growth within the Millennium-II simulation box of 100Mpc/$h$ size \citep{Kim2012b}. As can be seen in Fig.~\ref{DODED256Only}, a box of 100Mpc/$h$ size may not be large enough as ionized features can fill a significant fraction of the simulation volume, even at a mean mass averaged neutral hydrogen fraction of 0.45. 

In this section we describe a method to predict the 21-cm intensity map during reionization within larger volumes. The simulations we use for this include the Millennium \citep{Springel2005}, the GiggleZ \citep{Poole2015} and the Millennium-XXL \citep[MXXL,][]{MXXL} simulations. These large volumes are required to model forthcoming 21cm simulations. Note that we rescale the dark matter density distributions of the Millennium, MXXL and GiggleZ-main simulations to match the Millennium-II simulation in order to avoid different results caused by different redshift outputs (between z=7.272 and z=7.33) or different cosmologies. {This rescaling is necessary because different output redshifts or different cosmologies lead a deviation in the distribution width of dark matter overdensities. We adjust for this deviation by adding a multiplicitive factor to the logarithm of each density contrast (e.g., $\sim$1.1$\times$log(1+${\rm \delta_{dm,GiggleZ}}$)=log(1+${\rm \delta_{dm,Millennium-II}}$)}). A summary of dark matter simulations is given in Table~\ref{Nbodies}.
%\footnote{The cosmological
%parameters adopted for the Millennium Simulation \citep{Springel2005} are the same as for the Millennium-II simulation. The Millennium simulation has 125 times the volume of the Millennium-II simulation with the same number of particles.}, 
% the GiggleZ \citep{Poole2015}%\footnote{The cosmological
%parameters adopted for the GiggleZ-main Simulation are the WMAP5 results. The GiggleZ-main simulation has 1000 times the volume of the Millennium-II simulation with same number of particles.}
 % and the Millennium-XXL \citep[MXXL,][]{MXXL}%\footnote{The cosmological
%parameters adopted for the MXXL simulation are the same as for the Millennium \& Millennium-II simulations. MXXL simulation has 216 times the volume of the Millennium simulation with $\sim$ 30 times more particles.} 
%simulations. 

\begin{table*}
\caption{
Some basic properties of the dark matter simulations used in the paper. L$_{\rm box}$ is the side length of the simulation box, N$_{\rm p}$ is the total number of simulation particles used, and $\epsilon$ is the Plummer-equivalent force softening of the simulation, in comoving units. m$_{\rm p}$ gives the mass of each simulation particle.}
\label{Nbodies}
\begin{tabular}{lccccc}
\hline
\hline
 & L$_{\rm box}$ [Mpc/$h$] & N$_{\rm p}$& $\epsilon$ [kpc/$h$] & m$_{\rm p}$ [M$_{\odot}$/$h$] & cosmology\\
\hline
Millennium-II & 100  & 10,077,696,000  & 1.0 & 6.89$\times$10$^{6}$ & Millennium \\
Millennium & 500 & 10,077,696,000  & 5.0   & 8.61$\times$10$^{8}$ & Millennium\\
Millennium-XXL & 3000 & 303,464,448,000 & 10.0 & 6.17$\times$10$^{9}$ &Millennium\\
GiggleZ-main & 1000 &10,077,696,000 & 9.3 & 7.52$\times$10$^{9}$&WMAP5\\
\hline
\end{tabular}
\end{table*}

\subsection{Monte Carlo realization of the $Q_{\rm cell}$ values within dark matter simulations}\label{MCmodels}
\begin{figure*}
\begin{center}
\includegraphics[width=7.8cm]{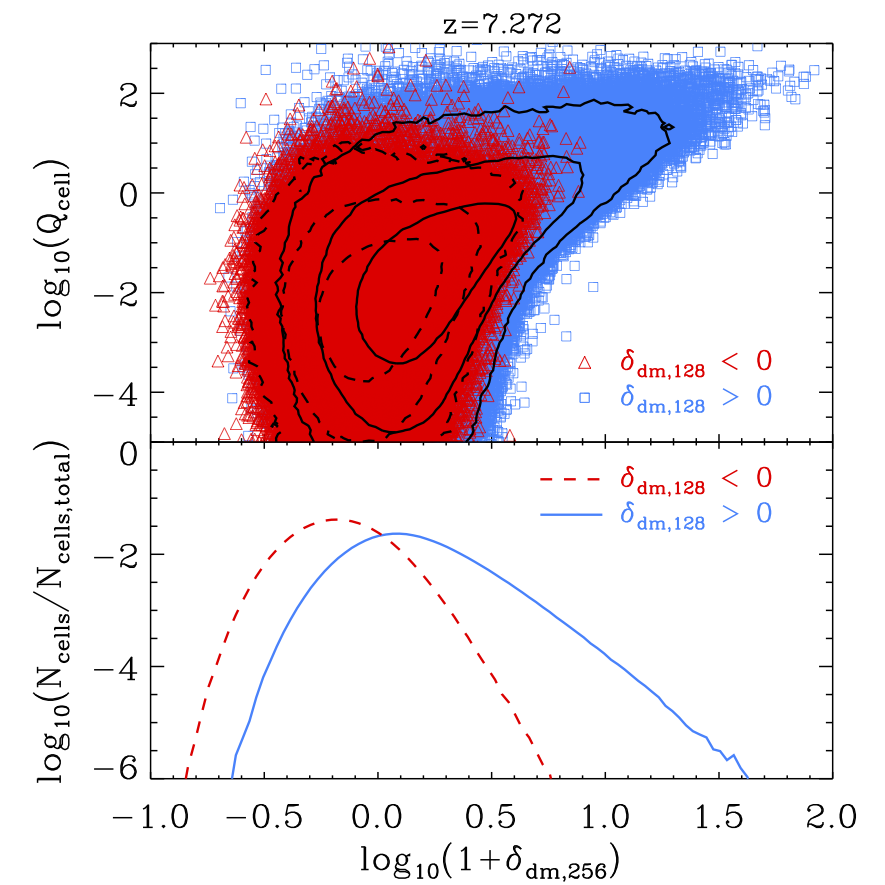}
\includegraphics[width=7.5cm]{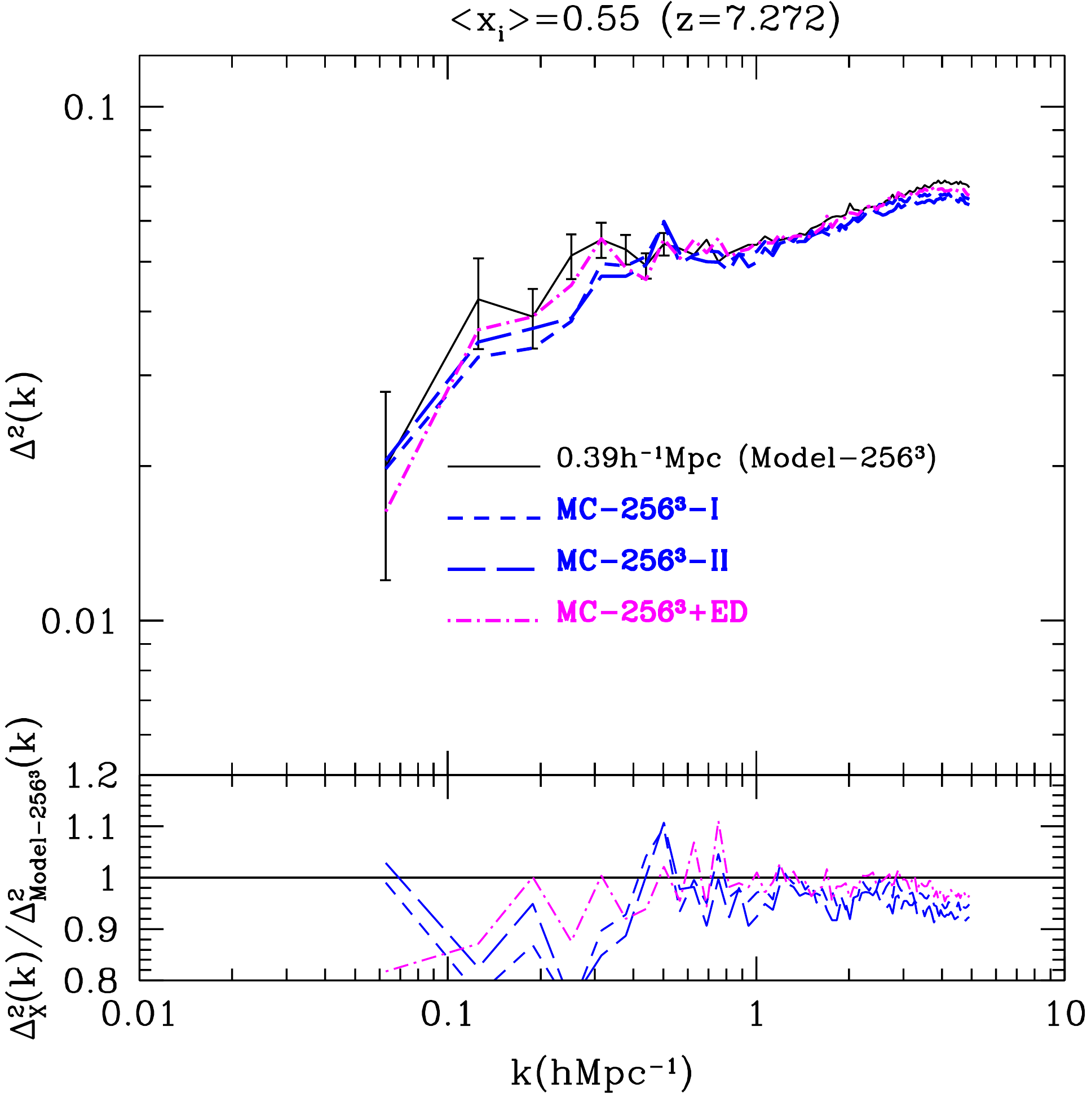}
\includegraphics[width=7.8cm]{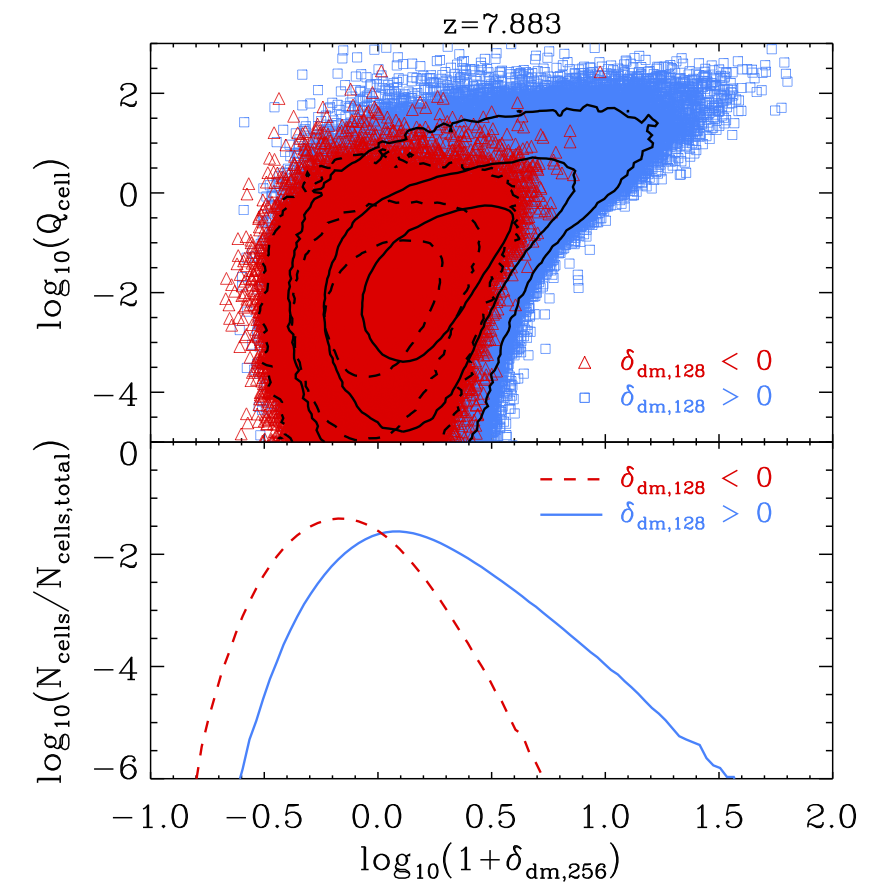}
\includegraphics[width=7.5cm]{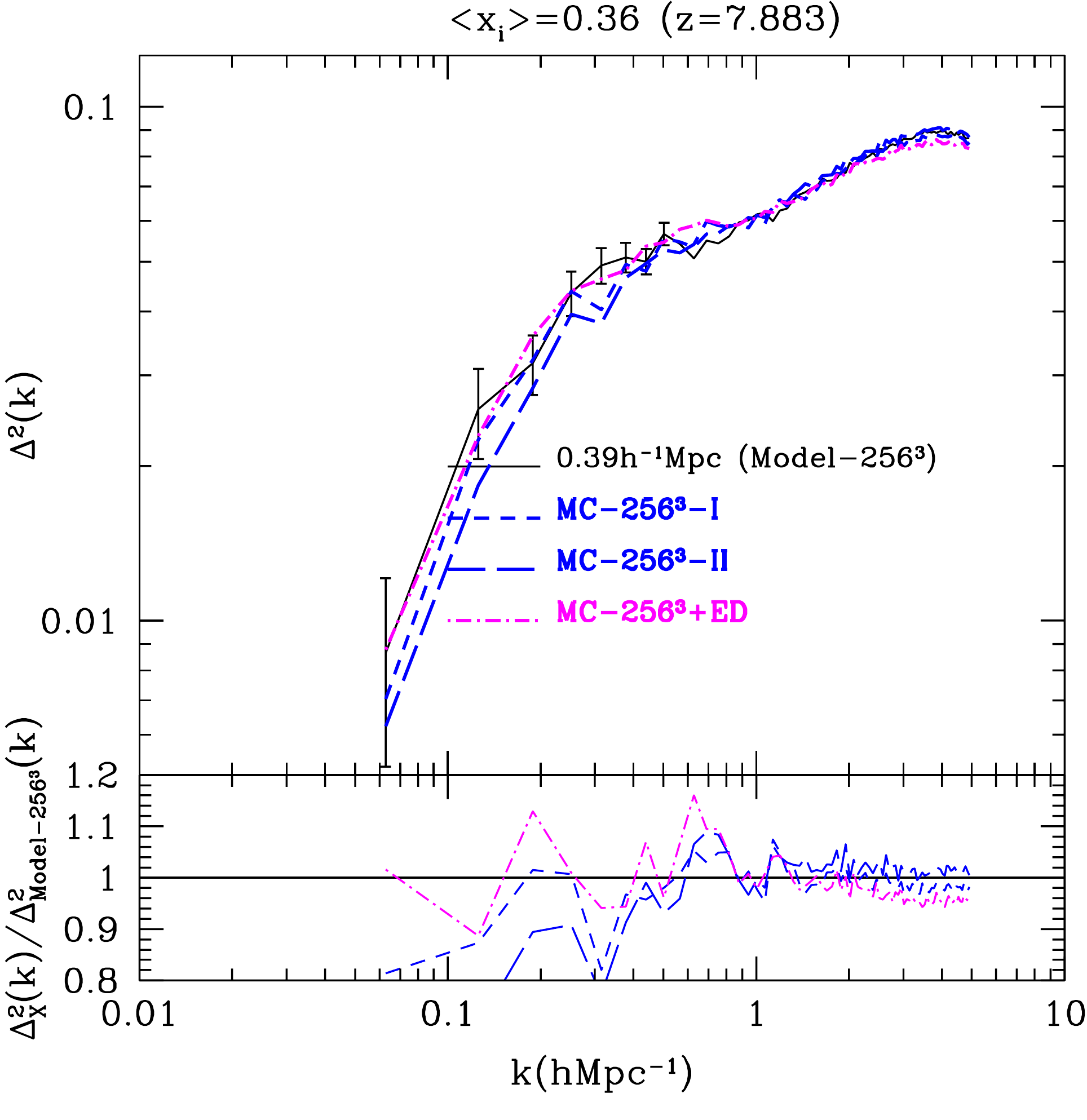}
\end{center}
\caption{Left panels show the distribution of Q values (top sub-panels) as a function of dark matter overdensity and number distributions of dark matter overdensities (bottom sub-panels) in the Model-256$^{3}$ simulation at two redshifts ($z$=7.272 (top), 7.883 (bottom)). The blue squares and red triangles correspond to over and under dense regions on large scale. The solid (dashed) line contours in left panels show 68.3\%, 95.4\%, and 99.7\% of this distribution for over (under) dense region. Right panels show the 21-cm power spectrum predictions using the Model-256$^{3}$, MC-256$^{3}$ models (blue lines) and MC-256$^{3}$+ED model (magenta line) for two redshifts. The fractional difference relative to the 256$^{3}$ model power spectrum is shown in the lower sub panel. 
} \label{DOD256}
\end{figure*}

{Before discussing application to large volumes we develop our method within the Millennium-II simulation, allowing us to test for systematics and errors in the method.}
We extract the $Q_{\rm cell}$ distribution of values (from {\bf eq.~\ref{Qvalue}}) as a function of dark matter overdensity (from the Millennium II dark matter simulation) using the luminosities from the GALFORM galaxy formation model. We refer to this default model as the Model-256$^{3}$ and to this distribution as the {\bf Q}value {\bf D}ark matter overdensity {\bf O}ccupation {\bf D}istribution; {\bf QDOD}. The top (bottom)-left panel of Fig.~\ref{DOD256} shows the distribution of $Q_{\rm cell}$ values as a function of dark matter overdensity for all pixels in the Model-256$^{3}$ model at z=7.272 (7.883). %Red triangles and blue squares correspond to overdense and underdense pixels respectively. 
%The red, orange, and yellow colour contours represent the regions enclosing 68.3\%, 95.4\%, and 99.7\% of this distribution, respectively.

The physics of galaxy formation produces a complex, nonlinear relation between the dark matter overdensity and $Q_{\rm cell}$ values. To populate the distribution of $Q_{\rm cell}$ values as a function of dark matter overdensity, we have binned by dark matter overdensity and measured the probability distribution of $Q_{\rm cell}$ values in each overdensity bin, $P\left[Q_{\rm cell}|(1+\delta_{\rm dm,cell})\right]$. To calculate the reionization structure within a large volume from the relation between the dark matter overdensity and $Q_{\rm cell}$ values, we then use a Monte-Carlo technique to populate the dark matter simulation (smoothed on the spatial scale of the cells in the reionization simulation) with $Q_{\rm cell}$ values from this distribution.\footnote{For comparison, \cite{Battaglia2013} reconstruct the ionization field using the best fit parametric form obtained from the radiation-hydrodynamic high resolution simulation that describes the correlation between the reionization redshift and dark matter overdensity field as a function of redshift.} {\bf {We calculate the $N_{\rm \gamma, cell}$ and $N_{\rm H, cell}$ using eqs. 4 and 5 based on the populated $Q_{\rm cell}$ values and $\delta_{\rm dm,cell}$ in a large volume simulation.} We follow the semi-numerical scheme as described in \S\ref{scheme} to find the ionization structure. In order to find the extent of ionized regions we therefore filter the resulting $N_{\rm \gamma, cell}$ and $N_{\rm H, cell}$ fields using a sequence of real-space top hat filters of radius R (from the cell size to box size), producing one smoothed ionization  field $Q_{\rm R}$ per radius using eq.~\ref{QvalueR}. We find the largest R for which the smoothed ionization field is greater than unity (i.e. ionized with $Q_{\rm R}$$>$1). All cells within radius R around
this point are considered ionized. We then account the overlap region of adjacent HII bubbles as in \S\ref{scheme} to achieve photon conservation.}
%{\bf We calculate the $N_{\rm \gamma, cell}$ and $N_{\rm H, cell}$ using eqs.~\ref{Qvalue} and \ref{nHI} based on the populated $Q_{\rm cell}$ values and $\delta_{\rm dm,cell}$ in a large volume simulation.} The extent of ionized regions is found by filtering the resulting $Q_{\rm cell}$ field using a sequence of real space filters {\bf (using eq.~\ref{QvalueR}) and then taking account the overlap region of HII bubbles} {as described in \S\ref{scheme}}.

\begin{figure*}
\begin{center}
\includegraphics[width=8.5cm]{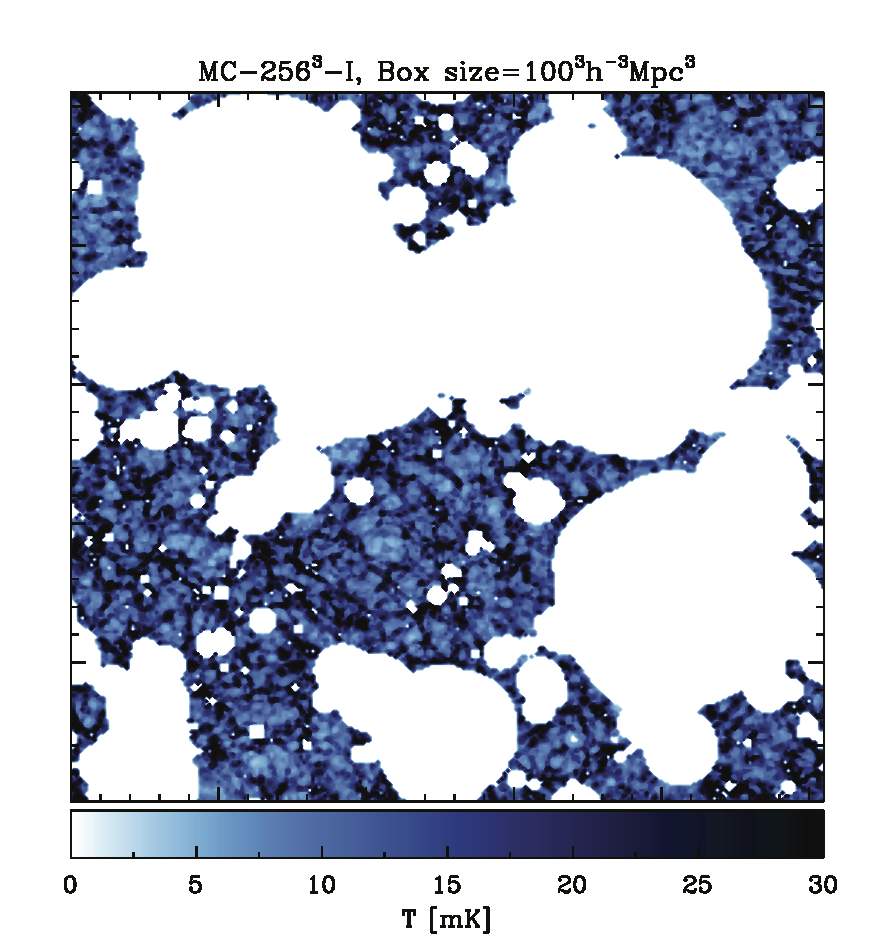}
\includegraphics[width=8.5cm]{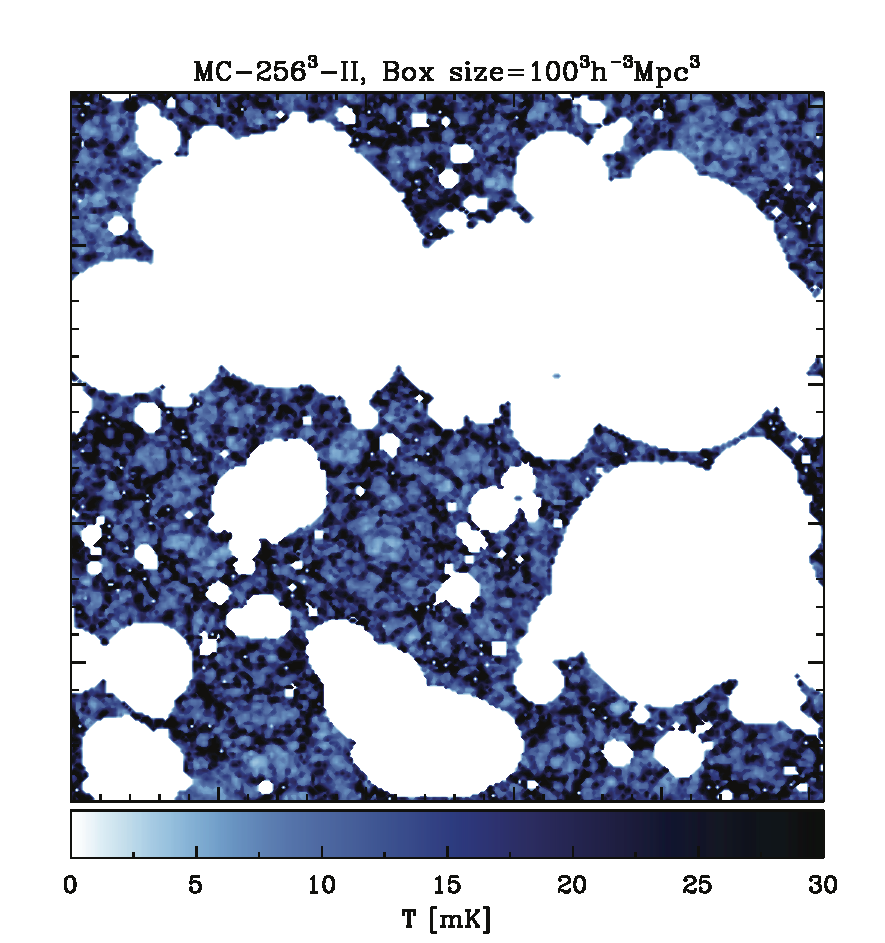}
\end{center}
\caption{Realisations {of 21cm intensity maps} of the MC-256$^{3}$ (cell size 0.39Mpc/$h$) models {at $z \sim$7.272 ($\left<x_{\rm i} \right>$$\sim$0.55)} with slices that are 0.39Mpc/$h$ deep. The colour shading shows the 21cm intensity in temperature units, as indicated by the bar.
} \label{DODED256MC}
\end{figure*} 

We note that this approach does not capture the possible correlation of ionization luminosities for cells separated by a distance $r$, and so may introduce noise into the ionization map due to the random assignment of ionizations at fixed $\delta_{dm,cell}$. However, we show that the effect of this on the power spectrum is negligible over large volumes, although the ionization field does show small differences on small scales (see the GiggleZ-500$^{3}$ models in Fig.~\ref{PSALL}). Moreover, on large scales the method does capture the very large scale clustering of ionising radiation in the linear regime, because the clustering of overdensities is described by the large volume N-body simulation.

To test our method, we show the {resulting ionization maps in Fig.~\ref{DODED256MC}} from two Monte-Carlo models calculated within the Millennium-II dark matter simulation (hereafter MC-256$^{3}$-I and II) on which the default Model-256$^{3}$ was based. {We also show the corresponding 21cm power spectrum} in the top and bottom right panels of Fig.~\ref{DOD256} for $z$=7.272 and 7.883. 
The right panels of Fig.~\ref{DOD256} show that the amplitudes and overall shapes of the 21cm power spectra from the MC-256$^{3}$ realisations are in reasonable agreement with Model-256$^{3}$. However, the amplitudes of 21cm power spectra from MC-256$^{3}$ models are $\sim$ 10\% lower than the Model-256$^{3}$ at large scales for both redshifts (see the ratio of MC-256$^{3}$ models to the Model-256$^{3}$ in bottom sub-panels of right panels of Fig.~\ref{DOD256}.)

\subsection{Environmental dependence on $Q_{\rm cell}$}
To improve the calculation, we note that $Q_{\rm cell}$ is related to not only dark matter overdensity but also the environment of dark matter overdensity. {We therefore choose a larger cell ($\sim$ $\times$ 8 in volume) surrounding the point containing the value of $Q_{\rm cell}$ to include any environmental effects. We summarise the cell size of models and the environmental cell size of models including the environmental effect in Table~\ref{Sizepixel}.} The left panels of Fig.~\ref{DOD256} show the distribution of $Q_{\rm cell}$ values (top-sub panels) in regions of over (blue squares)- and under (red triangles) density within {a 128$^{3}$ grid ($\delta_{\rm dm,128}$)} at z=7.272 and z=7.883. The solid (dashed) line contours in the sub panels of Fig.~\ref{DOD256} enclose 68.3\%, 95.4\%, and 99.7\% of this distribution for over (under) dense regions, respectively. $Q_{\rm cell}$ values on the 256$^{3}$ grids in the high overdensity group have statistically larger values than those in the low overdensity group. {We incorporated both conditional probabilities for $Q_{\rm cell}$ ($P\left[Q_{\rm cell}|(1+\delta_{\rm dm,256})|(\delta_{\rm dm,128}>0)\right]$ and $P\left[Q_{\rm cell}|(1+\delta_{\rm dm,256})|(\delta_{\rm dm,128}<0)\right]$) into our realisations.} The realisation including this large scale environmental dependence better matches the amplitude of the  model 21cm power spectrum at scales {between $k \sim$0.1$h$/Mpc and $k \sim$ 0.5$h$/Mpc} (MC-256$^{3}$+ED in the right panels of Fig.~\ref{DOD256}). It is therefore important to include the environmental effect in the simulation. We include this large scale environment effect in all subsequent models for the paper.

\subsection{Dependence of cell size}\label{Bigcell}
Having tested the method, we next expand our calculations to larger volumes.
In order to do this it is convenient to increase the cell size. 
We have therefore smoothed the cell size of our default simulation within the Millennium-II to 2Mpc/$h$ rather than the 0.39 Mpc/$h$ used in Fig.~\ref{DOD256}. As a result we decrease the number of cells in the Millennium-II simulation from 256$^{3}$ to 50$^{3}$ cells. We refer to this as the Model-50$^{3}$ simulation.

\begin{figure*}
\begin{center}
\includegraphics[width=7.8cm]{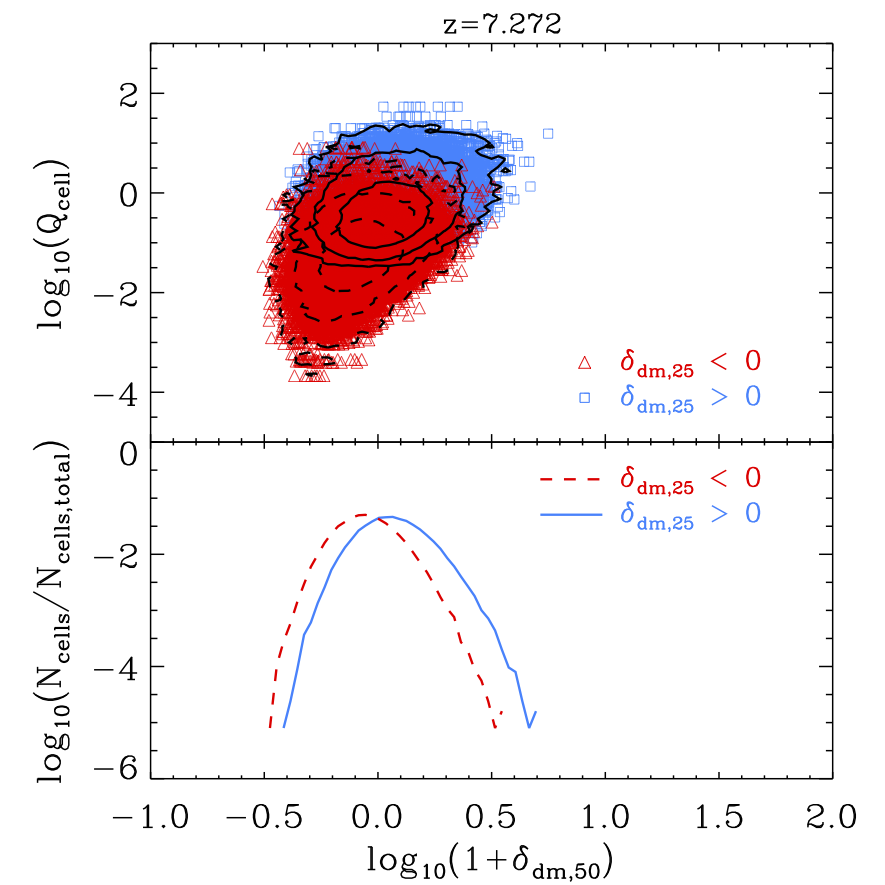}
\includegraphics[width=7.5cm]{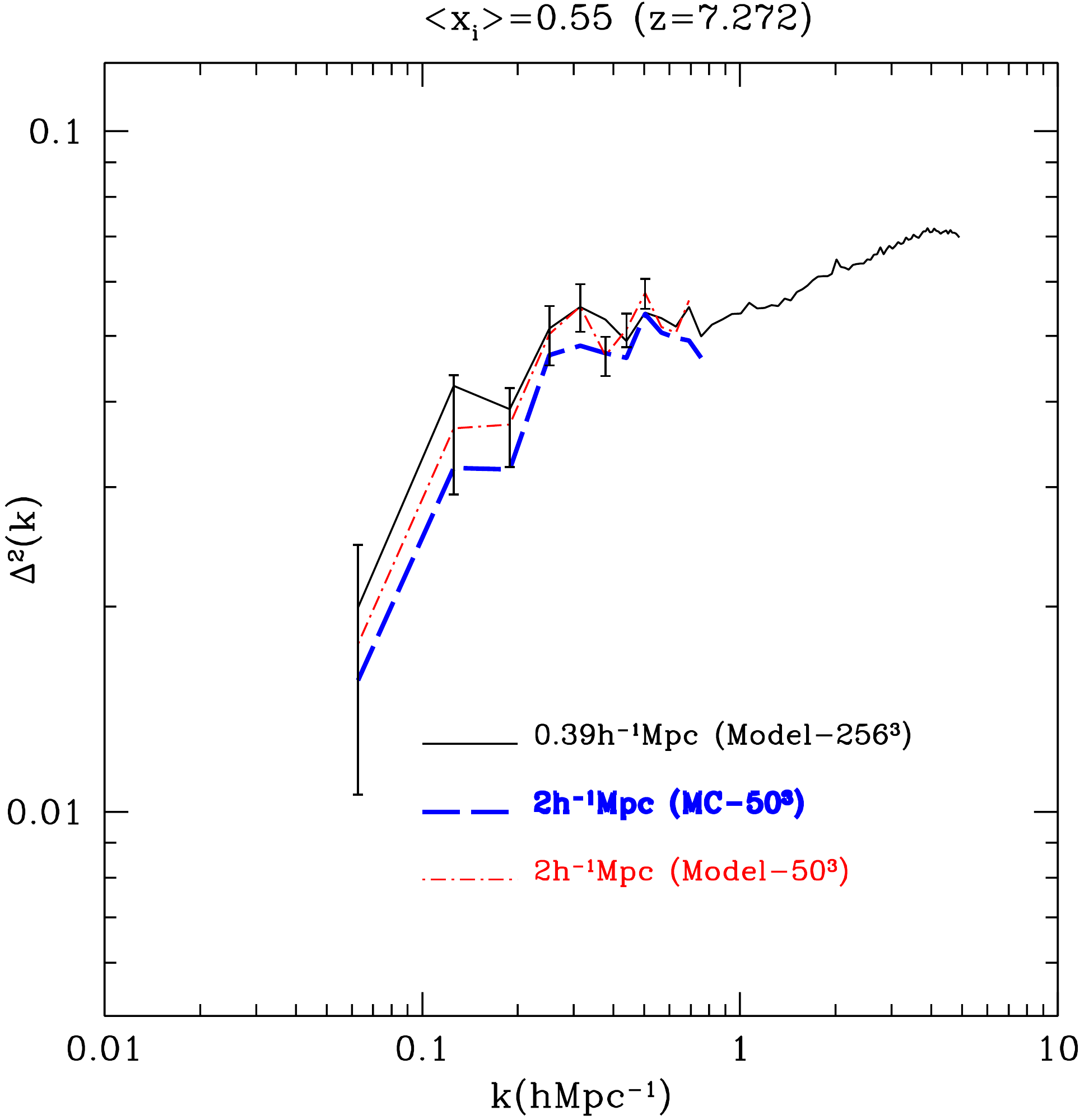}
\includegraphics[width=7.8cm]{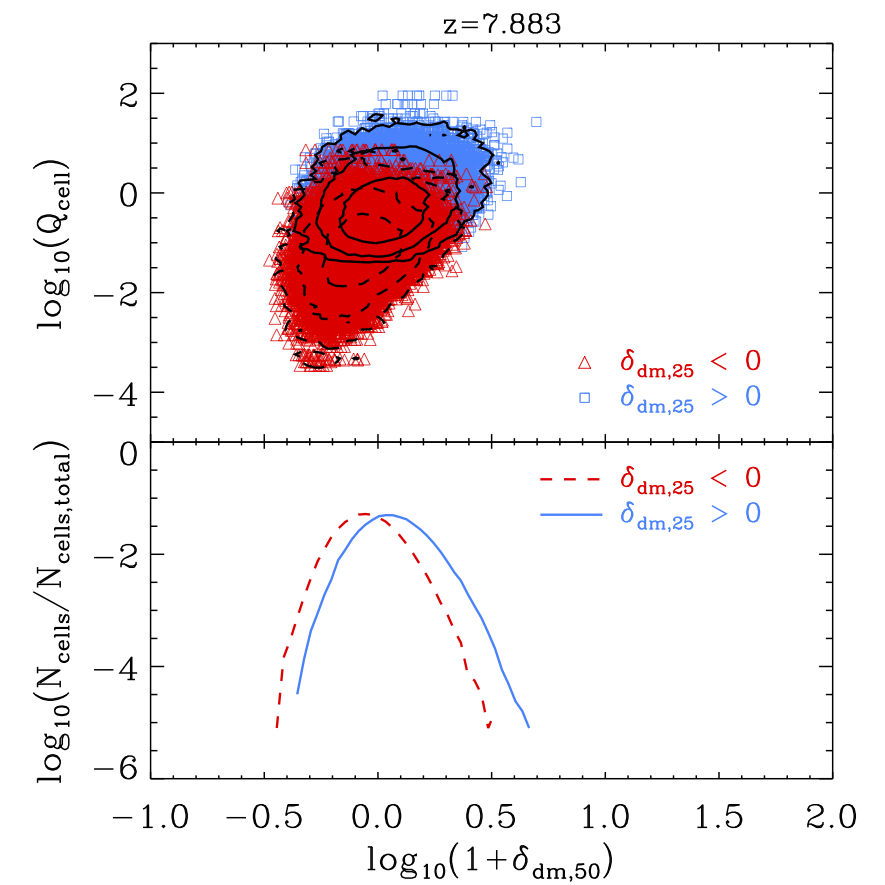}
\includegraphics[width=7.5cm]{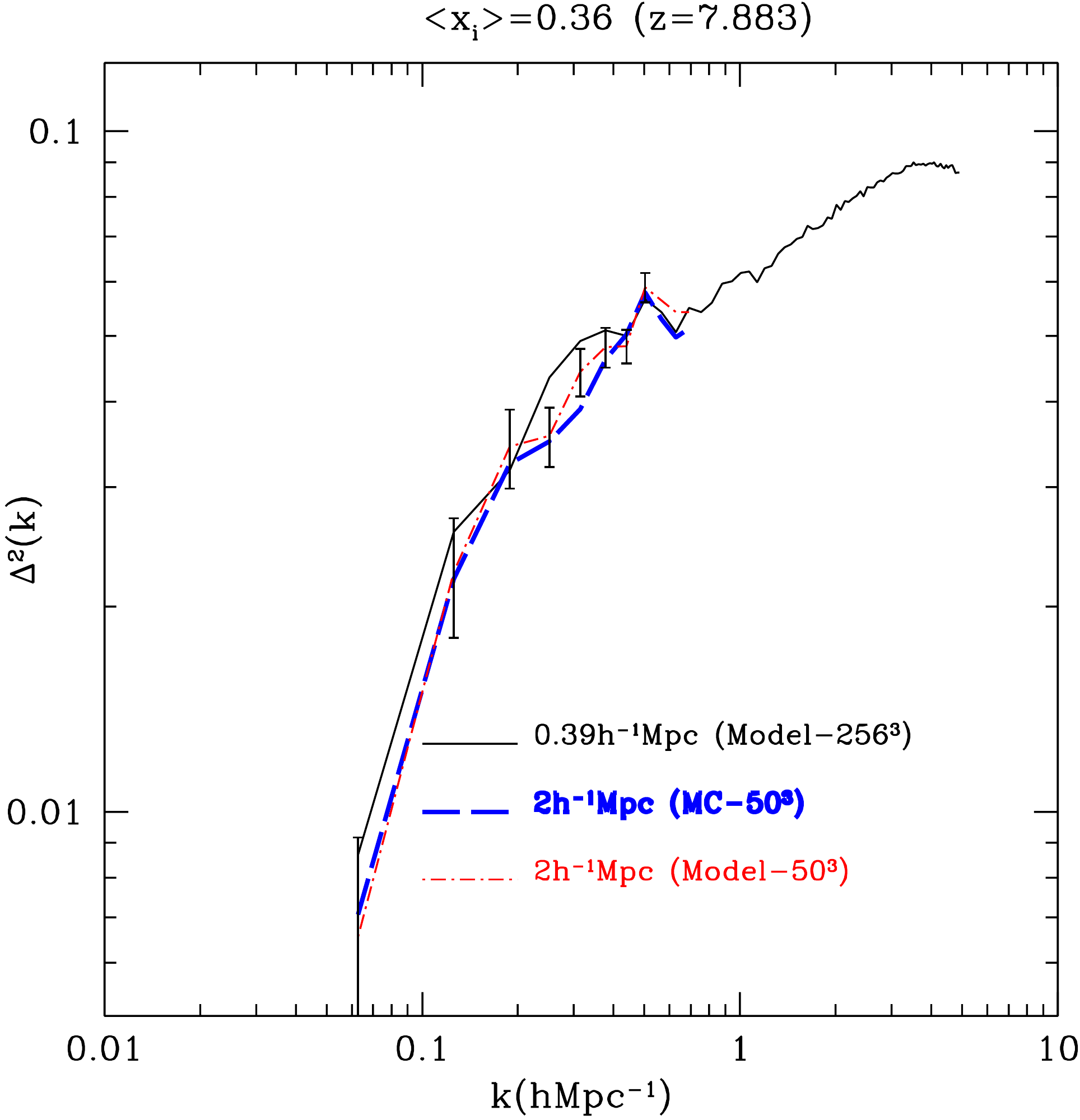}
\end{center}
\caption{Left panels show the distribution of Q value (top sub-panels) as a function of dark matter overdensity and number distributions of dark matter overdensities (bottom sub-panels) in the Model-50$^{3}$ simulation a 2Mpc/$h$ cell size at z=7.272 and 7.883. The blue squares and red triangles correspond to over and under dense regions. The solid (dashed) line contours in left panels show 68.3\%, 95.4\%, and 99.7\% of this distribution for over (under) dense region on large scale. Right panels show the 21-cm power spectrum predictions by the Model-50$^{3}$ and MC-50$^{3}$ simulations with the Model-256$^{3}$ simulation for comparison at two redshifts.}
\label{DOD50}
\end{figure*}

\begin{figure*}
\begin{center}
\includegraphics[width=8.5cm]{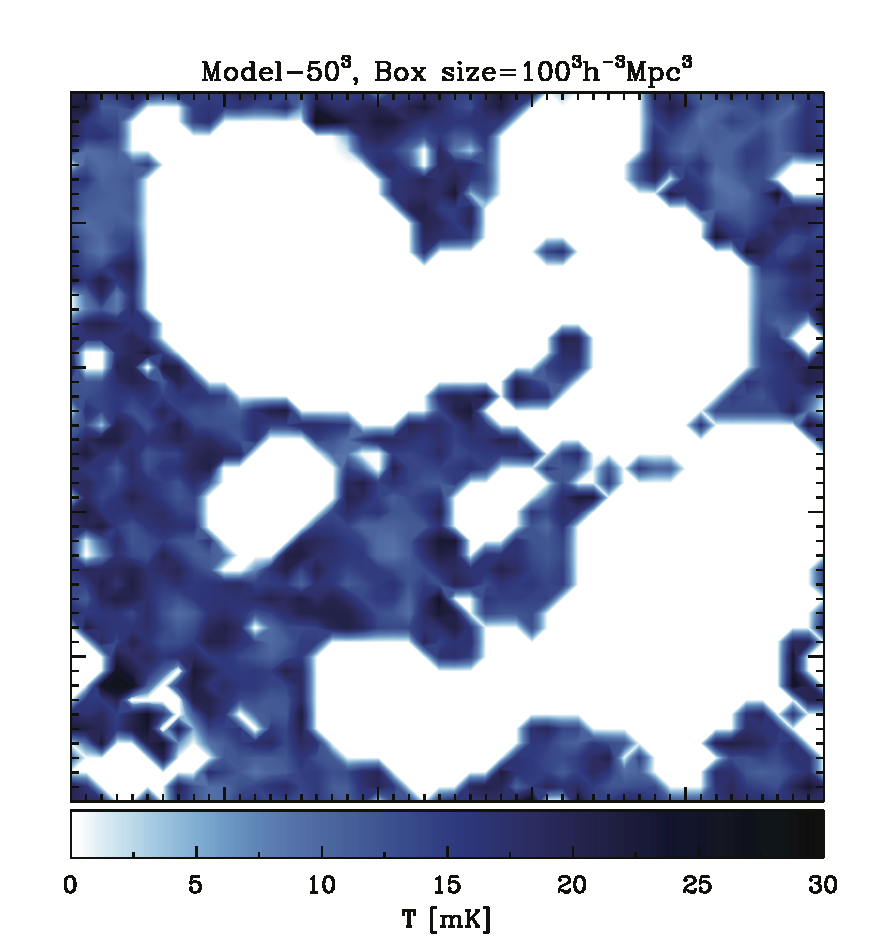}
\includegraphics[width=8.5cm]{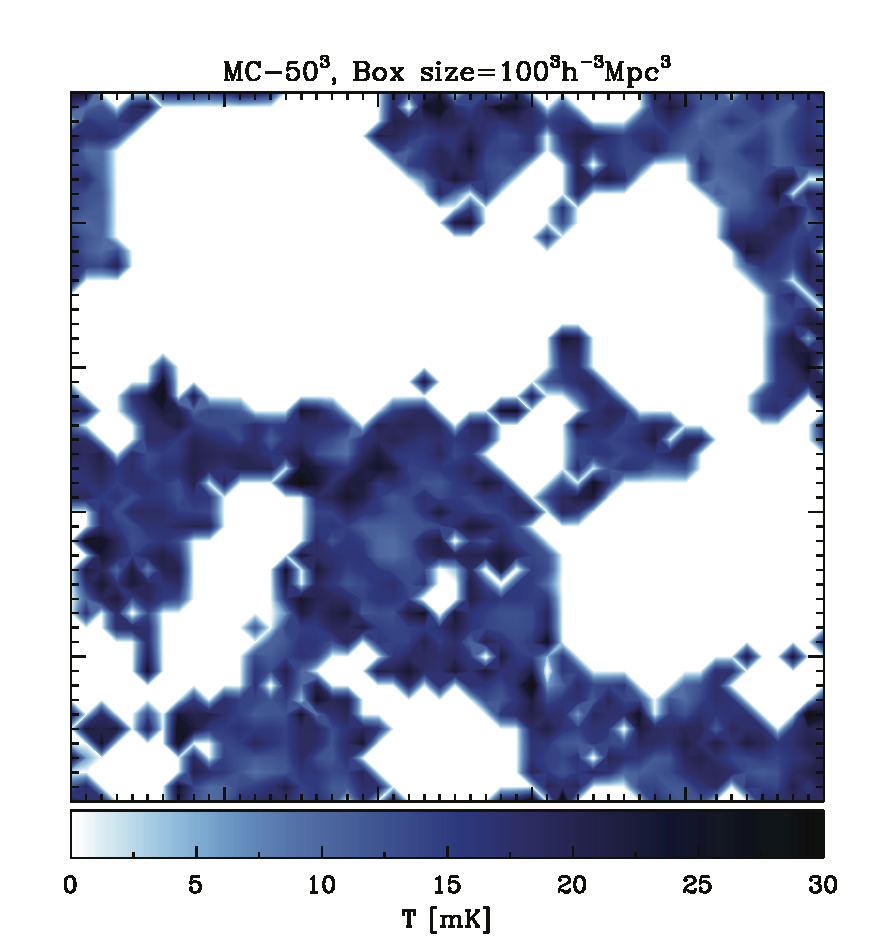}
\end{center}
\caption{The 21cm intensity maps of the Model-50$^{3}$ and the MC-50$^{3}$ models {at $z \sim$7.272 ($\left<x_{\rm i} \right>$
$\sim$0.55)} with cell size 2Mpc/$h$ in Sec.~\ref{Bigcell} and 2Mpc/$h$ deep. The colour shading shows the 21cm intensity in temperature units, as indicated by the bar.
} \label{DODED50}
\end{figure*}

Fig.~\ref{DOD50} shows results for this lower resolution that correspond to those in Fig.~\ref{DOD256} for the $Q_{\rm cell}$ value distribution as a function of overdensity (with environment effect; i.e., red triangles and blue squares). We see that the $Q_{\rm cell}$ value distribution from the Model-50$^{3}$ model has a much tighter relation than in the Model-256$^{3}$ model both z=7.272 and 7.883, as a result of smoothing on the larger grid. The solid (dashed) line contours in the left panels of Fig.~\ref{DOD50} show 68.3\%, 95.4\%, and 99.7\% of this distribution for over (under) dense regions, respectively. {We use this QDOD as described in \S\ref{MCmodels} to calculate Monte-Carlo realizations of the ionization structure on a 50$^{3}$ grid. Two examples are shown in Fig.~\ref{DODED50}.} The corresponding redshifted 21-cm power spectra from these two models are noisy, but again show good agreement (see Fig.~\ref{DOD50}). For comparison, we also show the power spectrum from the Model-256$^{3}$. Importantly the agreement between the Model-50$^{3}$ and the Model-256$^{3}$ power spectra is good. {These calculations provide a demonstration that our method for constructing Monte-Carlo ionization fields within the parent volume of the reionization simulation produces accurate power spectra, and is insensitive to the grid resolution.}

\subsection{Application to larger volumes}\label{Largevolume}

%{Two examples (MC-256$^{3}$ models) of the application of QDOD to the Millennium-II dark matter simulation are shown in Fig.~\ref{DODED256MC} with different random {\bf realizations} for QDOD}. These demonstrate that the realisations of the ionization structure agree well with the default Model-256$^{3}$ simulation in the top left panel. Differences are apparent on small scales, but all the gross features larger than a few Mpc are consistent. This similarity is illustrated by the agreement of the power spectra between different realizations in top right panel of Fig.~\ref{DOD256}. In Fig.~\ref{DODED50}, we show the 21cm intensity maps from the Model-50$^{3}$ and MC-50$^{3}$ simulations for comparison to 256$^{3}$ models. 

%\subsection{Large volume simulations}\label{Largevolume}

We next apply our method for generating 21-cm intensity maps to the Millennium and the GiggleZ-main simulations. As above, we generate $Q_{\rm cell}$ values in the Model-50$^{3}$ model which is based on the Millennium II dark matter simulation and includes the low mass galaxies that drive reionization. The Model-50$^{3}$ model has a cell size of 2Mpc/$h$. This cell size corresponds to a grid size of 250$^{3}$ cells in the Millennium simulation and 500$^{3}$ cells in the GiggleZ-main simulation (cf. see also \cite{Ahn2012} for sub-grid modelling).
A summary of models is given in Table~\ref{Sizepixel}. 

\begin{figure*}
\begin{center}
\includegraphics[width=1.cm]{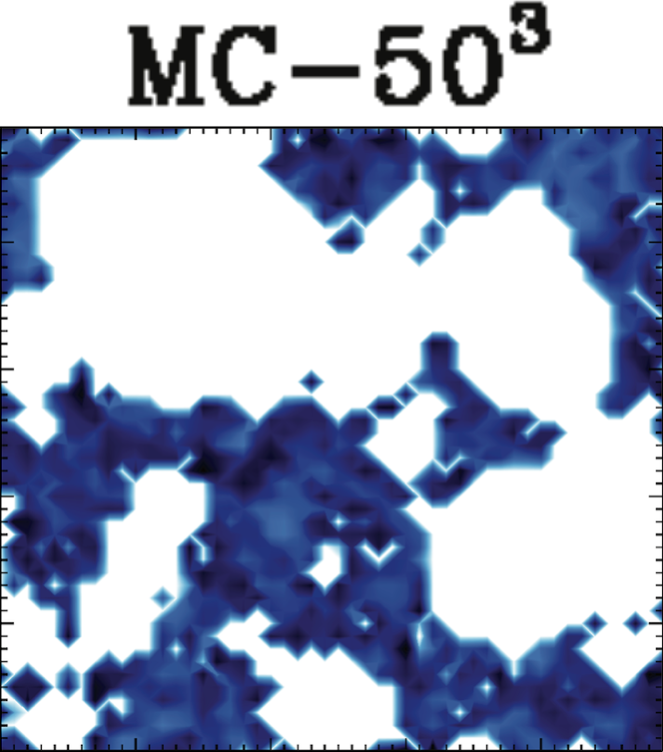}
\includegraphics[width=5.cm]{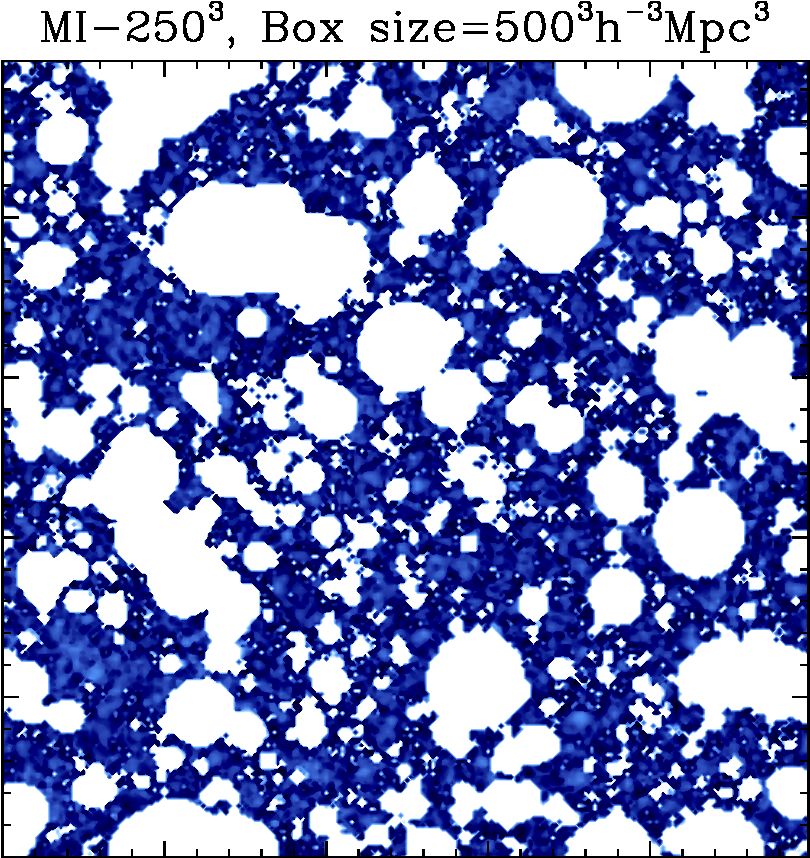}
\includegraphics[width=10.cm]{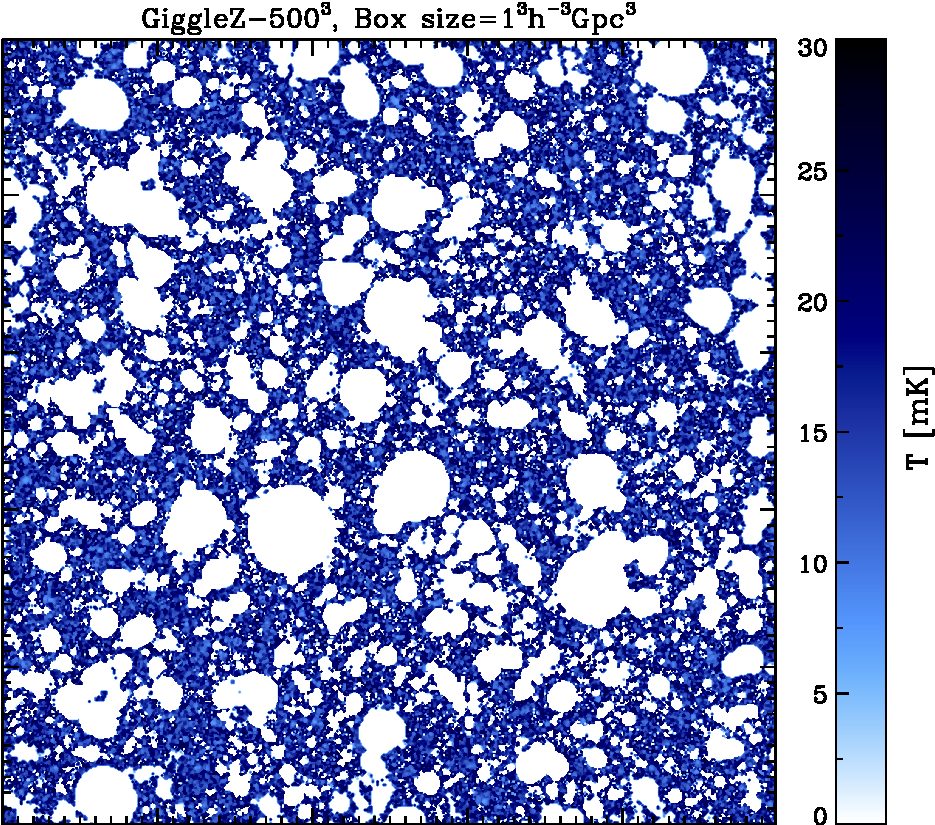}
\end{center}
\caption{The three panels show the 21-cm intensity maps from the MC-50$^{3}$ (100Mpc/$h$ box size), MI-250$^{3}$ (500Mpc/$h$), and GiggleZ-500$^{3}$ (1000Mpc/$h$) models {at $z \sim$7.272 ($\left<x_{\rm i} \right>$$\sim$0.55) with cell size 2Mpc/$h$}. All models use the Lagos11 model. The size of the figures correspond to the relative box size of simulations. The slices are 2Mpc/$h$ deep.} \label{HIDIFF}
\end{figure*}

\begin{figure*}
\begin{center}
\includegraphics[width=8.5cm]{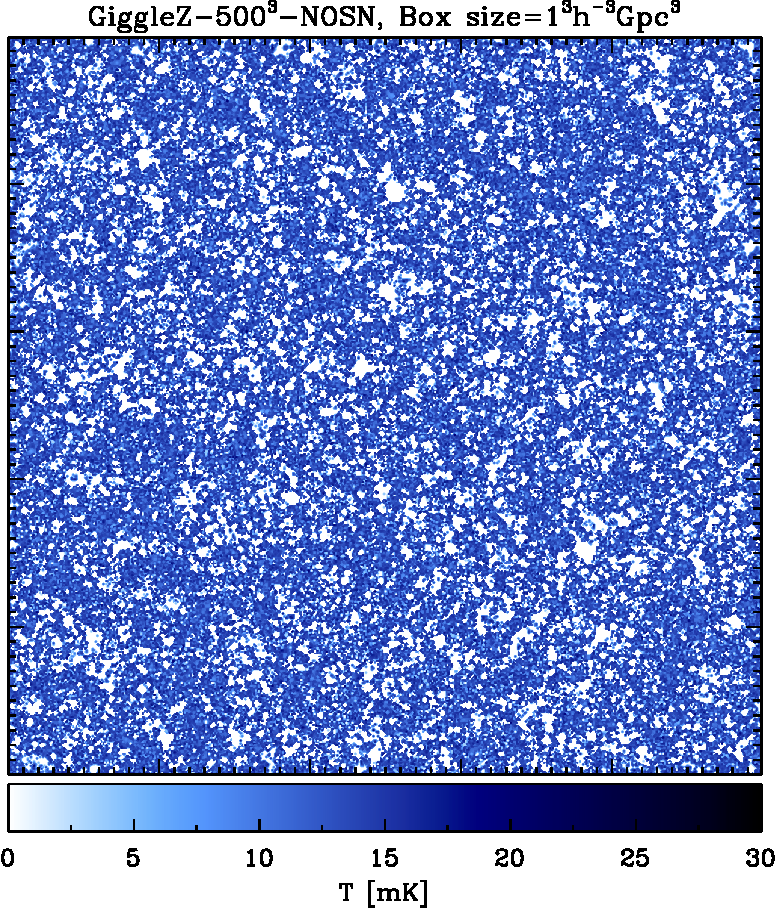}
\includegraphics[width=8.5cm]{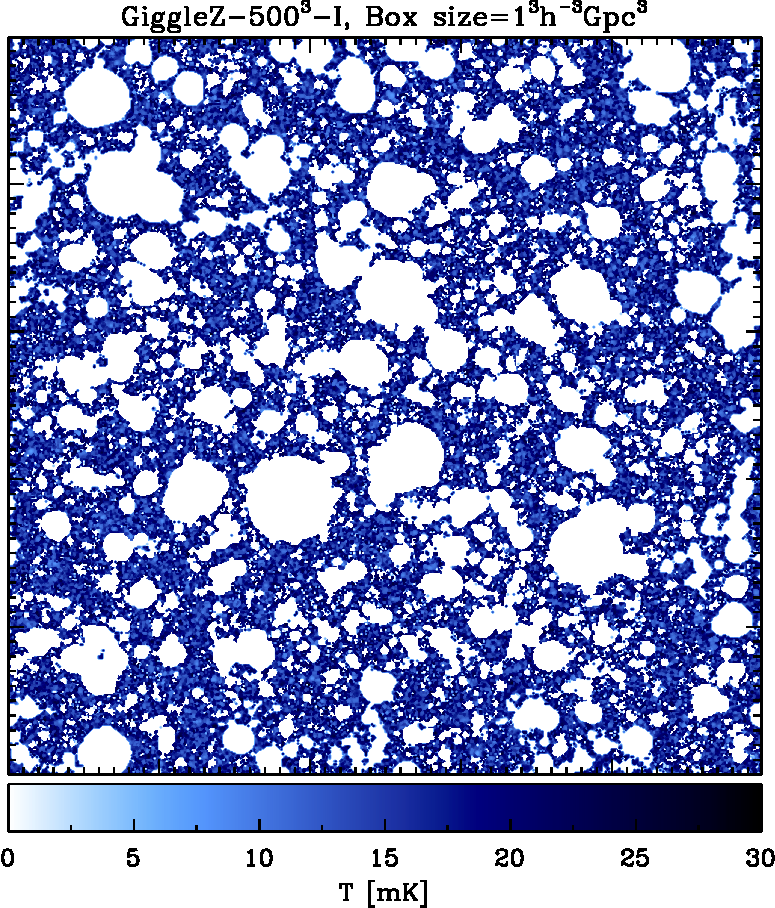}
\end{center}
\caption{The 21-cm intensity maps for the GiggleZ-500$^{3}$-NOSN (which use the NOSN galaxy formation model from \citet{Kim2013}) and GiggleZ-500$^{3}$-I (use the Lagos11 galaxy formation model) simulations {at $z \sim$7.272 ($\left<x_{\rm i} \right>$$\sim$0.55) with cell size 2Mpc/$h$}. The slices are 2Mpc/$h$ deep.} \label{GDOD}
\end{figure*}

\begin{figure}
\begin{center}
\includegraphics[width=8.5cm]{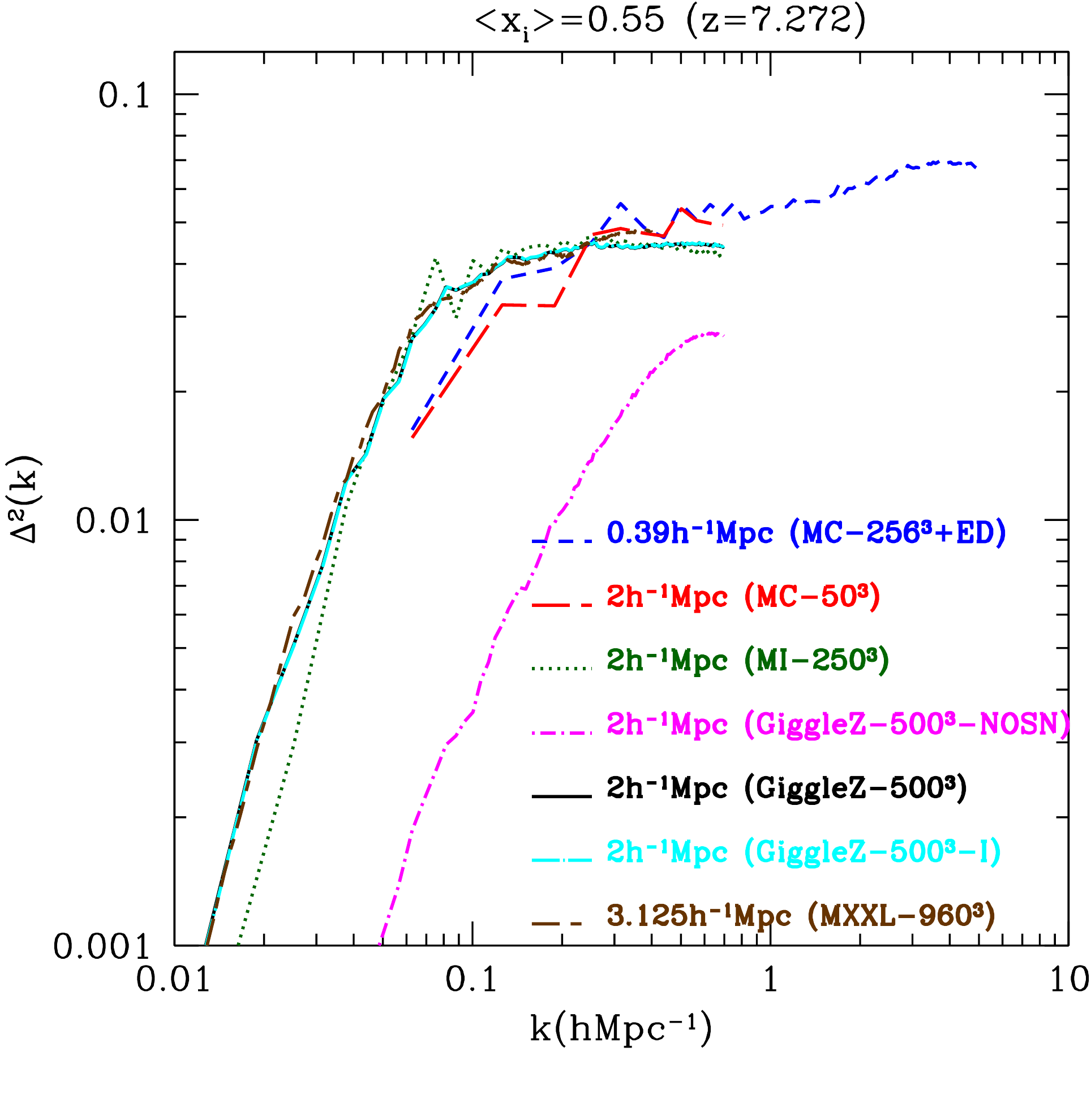}\vspace{-0.5cm}
\end{center}
\caption{21cm power spectrum predictions with comparison to the power spectrum from the high resolution simulation MC-256$^{3}$ model. The simulations are labelled in the figure.} \label{PSALL}
\end{figure}

Fig.~\ref{HIDIFF} shows the resulting reionization maps. The corresponding 21-cm power spectra for these models are shown in Fig.~\ref{PSALL}. The 21-cm power spectra from the models show good agreement for wavenumbers $k$ between 0.1$h$/Mpc and 1$h$/Mpc. However the larger 21-cm maps, from the Millennium and GiggleZ-main simulations, allow the 21-cm power spectrum to be extended to much larger scales. We also include a model that does not include supernovae feedback (hereafter GiggleZ-500$^{3}$-NOSN) based on the NOSN-0 model in \cite{Kim2013}. {For the NOSN model, we turn off feedback by supernovae in the default model, and change the free parameters (f$_{\rm esc}$ and F$_{c}$) to obtain $\left<x_{\rm i}\right>=0.55$}. The 21-cm map in the left panel of Fig.~\ref{GDOD} shows that the typical HII bubble size is much smaller than for models which include supernovae feedback. This is imprinted on the 21-cm power spectrum in Fig.~\ref{PSALL} which shows that the amplitude of the 21-cm power spectrum for the GiggleZ-500$^{3}$-NOSN model is much lower than the default model. {This is because the NOSN model has a much larger contribution to the ionizing photon budget from low mass haloes than the default model \citep[see more details in][]{Kim2013}. } 

To further test whether the Monte-Carlo method introduces power into the intensity distribution, we have generated {another random realisation within the GiggleZ-main simulation (hereafter the GiggleZ-500$^{3}$-I model)}. If the QDOD works correctly these two realisations should be statistically similar. The right panel of Fig.~\ref{GDOD} shows the resulting 21-cm map, with the corresponding  21-cm power spectra plotted in Fig.~\ref{PSALL}. Small scale differences can be seen by comparing the intensity maps for the GiggleZ-500$^{3}$ and GiggleZ-500$^{3}$-I. However, the power spectra are the same at the percent level across the full range of wavenumber $k$, indicating that the small differences seen in the power spectra shown in Fig.~\ref{DOD256} were due to the small volume rather than being due to stochasticity in the method.

\begin{figure*}
\begin{center}
\vspace{-0.5cm}
\includegraphics[width=18cm]{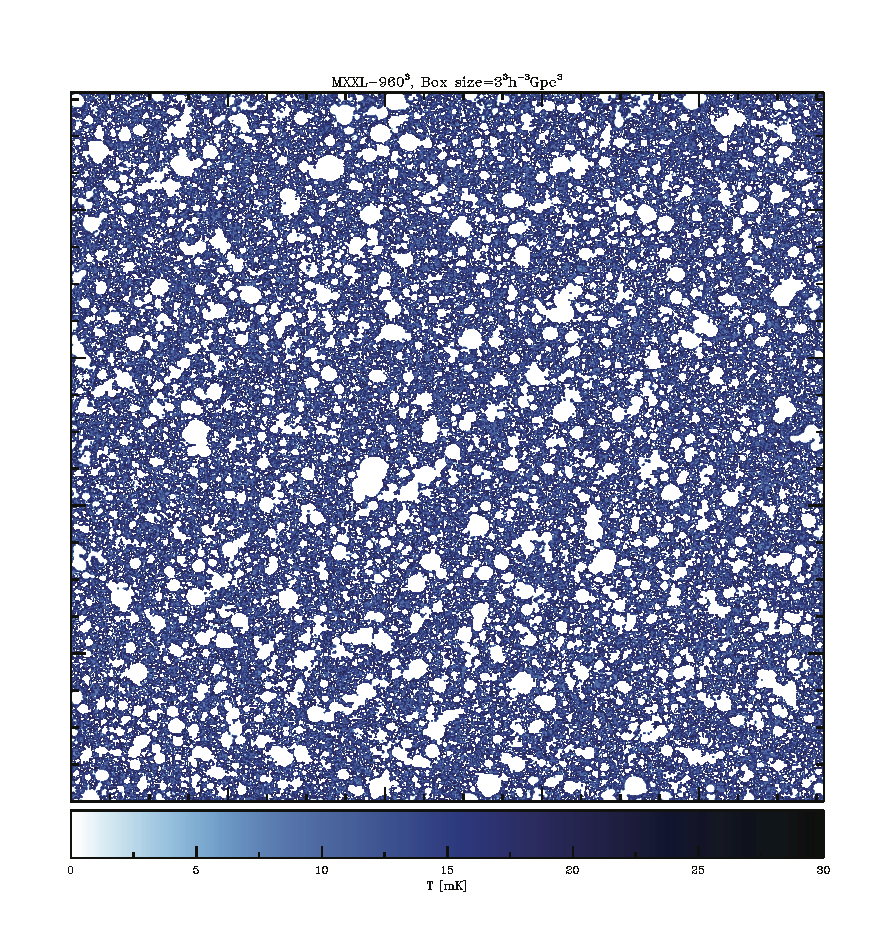}
\end{center}
\vspace{-1.2cm}
\caption{The 21-cm intensity map of the MXXL-960$^{3}$ simulation which has a box size of 3000Mpc/$h$ {at $z \sim$7.272 ($\left<x_{\rm i} \right>$$\sim$0.55) with cell size 3.125Mpc/$h$}. The slice is 3.125Mpc/$h$ deep.} \label{HIMAPALL}. 
\end{figure*}

We next apply the QDOD method to the Millennium XXL simulation which has a 3Gpc/$h$ box size (hereafter MXXL-960$^{3}$ model). We use the QDOD from the Model-256$^{3}$ model smoothed on a 3.125Mpc/$h$ (32$^{3}$ grids) to populate $Q_{\rm cell}$ values onto the Millennium XXL dark matter simulation (Fig.~\ref{HIMAPALL}). The simulated 21-cm power spectrum of these simulations is shown in the Fig.~\ref{PSALL}. We note that on large scales light-cone effects become important \cite[]{BattagliaIII}. %Comparison of our models to data would require construction of a light-cone as do all large volume simulations.

\begin{table*}
\caption{
The box size for N-body dark matter simulation we used models in this paper, the number of cells (number of cells to include environmental effect), and cell size (cell size to include environmental effect).}
\label{Sizepixel}
\begin{tabular}{ccccc}
\hline
Model & Box size& N-body simulation &$\#$ of cells&Cell size\\
\hline
\hline
MC-50$^{3}$ & 100Mpc/$h$& Millennium-II &50$^{3} $(25$^{3})$&2 (4) Mpc/$h$\\
\hline
MI-250$^{3}$ & 500Mpc/$h$& Millennium & 250$^{3}$ (125$^{3})$&2 (4) Mpc/$h$\\
\hline
HR-60$^{3}$ & 125Mpc/$h$ & GiggleZ-HR & 60$^{3}$ (32$^{3}$)&2.08 (4.16) Mpc/$h$\\
\hline
GiggleZ-500$^{3}$  &1000Mpc/$h$& GiggleZ-main &500$^{3}$ (250$^{3}$)&2 (4) Mpc/$h$\\
\hline
GiggleZ-500$^{3}{\rm -I}$  &1000Mpc/$h$& GiggleZ-main&500$^{3}$ (250$^{3}$)&2 (4) Mpc/$h$\\
\hline
GiggleZ-500$^{3}$-NOSN & 1000Mpc/$h$&GiggleZ-main&500$^{3}$ (250$^{3}$)&2 (4) Mpc/$h$\\
\hline
MXXL-960$^{3}$ & 3000Mpc/$h$&Millennium-XXL & 960$^{3}$ (750$^{3}$)& 3.125 (4) Mpc/$h$\\
\hline
\end{tabular}
\end{table*}

\section{21cm power spectrum predictions from large volume simulations}\label{Imp}
In this section, we use our simulations to {discuss} the effect of simulation volumes {on the large scale power spectrum} (\S\ref{ILV}). We also discuss the large scale 21cm power spectrum predictions from different star formation laws, and the presence of SNe feedback and photoionisation feedback (\S\ref{OI}).

\subsection{Large scale predictions of 21cm power spectrum}\label{ILV}
Here we investigate predictions for the 21cm power spectrum on the largest scales. \cite{iliev13} performed the largest numerical simulations of reionization to date, showing that large scale power continues to increase as volume increases, owing to the effect of large scale power on structure formation. 
We have used two sets of simulations, binned on a $\sim$2Mpc/$h$ (3.125Mpc/$h$ for the MXXL-960$^{3}$) scale, to investigate the effect of simulation volume on predictions for the 21cm power spectrum. One set includes the MC-50$^{3}$ model (100Mpc/$h$), the MI-250$^{3}$ model (500Mpc/$h$), and the MXXL-960$^{3}$ model (3Gpc/$h$) which are based on the Millennium simulation cosmology. The other {set} is the HR-60$^{3}$ model (a GiggleZ simulation which has 125Mpc/$h$ box size{; hereafter GiggleZ-HR}) and the GiggleZ-500$^{3}$ model (1000Mpc/$h$). The GiggleZ simulations are based on the WMAP7 cosmology.  

\begin{figure*}
\begin{center}\vspace{0cm}
\includegraphics[width=8cm]{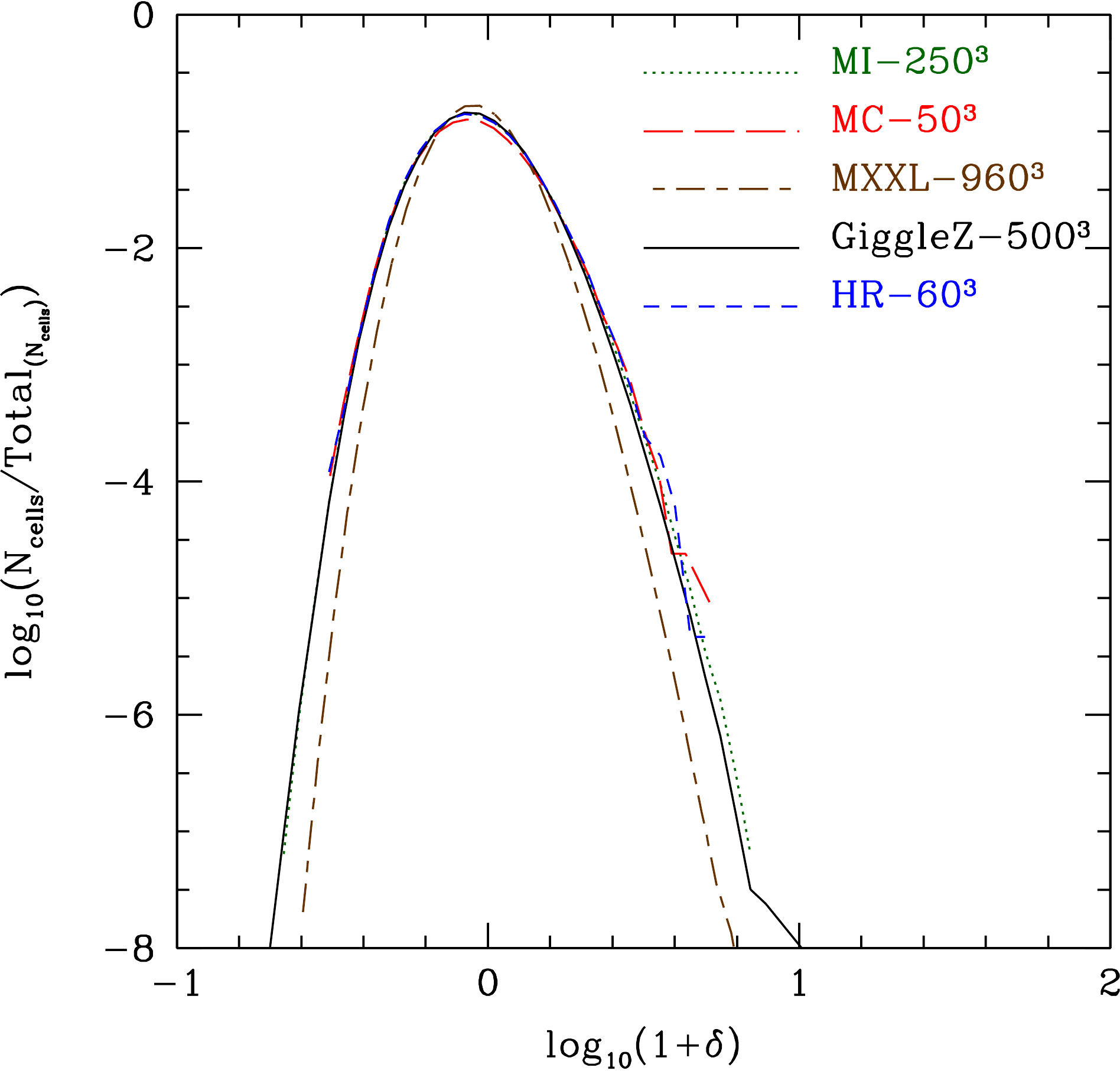}
\includegraphics[width=8cm]{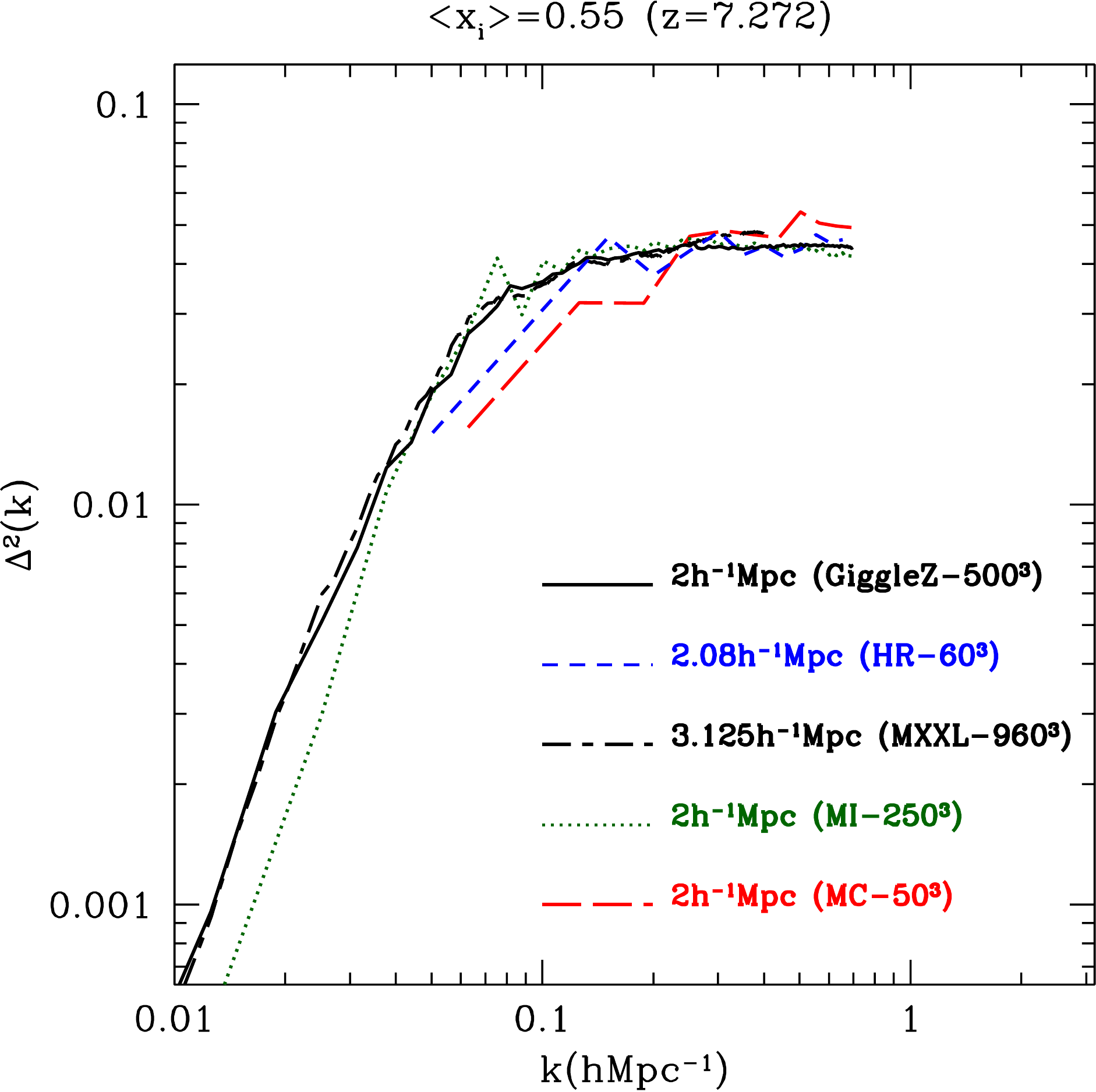}

\vspace{0cm}
\end{center}
\caption{The left panel show the number of cells as a function of dark matter overdensity and redshifted 21-cm power spectra from the set of MC-50$^{3}$, MI-250$^{3}$ (2Mpc/$h$ cell size), and MXXL-960$^{3}$ (3.125Mpc/$h$ cell size) models together with the set of  HR-60$^{3}$ (2.08Mpc/$h$ cell size)  and GiggleZ-500$^{3}$ (2Mpc/$h$ cell size) simulations. The right panel shows redshifted 21-cm power spectra of the models. Note that we plot rescaled dark matter number density distribution as described in \S\ref{Gmodel}} \label{DHODD}
\end{figure*}

The left hand panel of Fig.~\ref{DHODD} shows the distribution of dark matter overdensity for these models. The models show nearly identical distributions (note that the MXXL-960$^{3}$ has a narrower distribution than the other simulations because it is based on a 3.125Mpc/$h$ cell size). However, the relatively small box simulations (MC-50$^{3}$ and HR-60$^{3}$ models) have no overdensities greater than 4.5. The right hand panel shows the resulting 21cm power spectra. We see that there is significant extra power in the observational window {for} $k$$<$0.1$h$/Mpc within the (500Mpc/$h$)$^{3}$ {volumes of the Millennium, GiggleZ, and MXXL} than in the smaller (100Mpc/$h$)$^{3}$ simulation. {We find that the power spectra have converged at 0.01 $\leq k \leq$0.1 $h$/Mpc for volumes of (500Mpc/$h$)$^{3}$.} %However, the much larger MXXL simulation shows agreement with the 500Mpc/$h$ Millennium in the observational window around $k$$\sim$0.1$h$/Mpc, indicating convergence of reionization simulations with volume $>$ (500Mpc/$h$)$^{3}$. This indicates that simulation volumes of order 500Mpc/$h$ are sufficient to properly interpret the large scale features of reionzation \citep{iliev13}.
%{\bf However, we should be needed larger volumes to observe a convergence, which might be relevant towards the end of reionization. 
{However since larger bubbles form in the highly ionized stage of reionization, we may need even larger volume simulations to see the convergence of the predicted 21cm power spectrum at lower $z$.}

We note that the highest overdensity bins in MI-250$^{3}$, MXXL-960$^{3}$, and GiggleZ-500$^{3}$ models exceed values available in the input MC-50$^{3}$ model. However, these overdensities are very rare. To test the importance of these large over densities we put either $Q_{\rm cell}$=0 or $Q_{\rm cell}$ equal to the highest overdensity bin of MC-50$^{3}$. The predicted 21cm power spectra from these two different assumptions are nearly identical, indicating that these very rare and large overdensities do not contribute to the statistics of reionization. 
 
\subsection{Observational implications}\label{OI}
The first-generation low-frequency telescopes, such as MWA and LOFAR, aim to detect the slope and amplitude of the redshifted 21cm power spectrum \citep{lidz2008}. Following the analysis in \cite{Kim2012a} we calculate the slope and amplitude of the predicted redshifted 21cm power spectrum using large volume simulations of reionization. The simulations have a large enough volume to avoid the issue of sample variance near the central wavenumbers ($k$=0.2 and 0.4$h$/Mpc {corresponding to the point on the power spectrum at which {observables} will likely evaluate the amplitude and gradient from the MWA}). To quantify the effects of star formation law, we use implementations of GALFORM from \cite{Lagos2012} \& \cite{Bower2006} (Lagos11 .vs. Bow06)\footnote{Lagos11 extended GALFORM by modelling the splitting of cold gas in the ISM into its HI and H2 components and by linking star formation explicitly to the amount of H2 present in a galaxy.}. To quantify the effect of photoionisation feedback, we compare the NOSN(V$_{\rm cut}$=30km/s) (turn off the SNe feedback) .vs. NOSN (no suppression) (turn off both the SNe and the photoionisation feedbacks) models. We use models with and without photoionisation feedback (NOSN(Vcut=30km/s) .vs. NOSN (no suppression))\footnote{Photoionisation is predicted to have a dramatic impact on star formation in
low mass galaxies. In the standard implementation of {{GALFORM}}, the effect of photoionisation feedback induced by the epoch of reionization is modelled
by imposing a circular velocity cut off $V_{\rm cut}$=30km/s  on gas cooling at redshifts below the redshift corresponding to the end of reionization $z_{\rm cut}$=10. We turn off the photoionization feedback by setting $V_{\rm cut}$=0 (no suppression of gas cooling).}. Finally to quantify the effect of SNe feedback we compare the model from \cite{Bower2006}, with a modified model in the absence of SNe feedback (NOSN(Vcut=30km/s))\footnote{The default GALFORM model (e.g., Bow06 and Lagos11) parameterizes the SNe feedback mass loading efficiency as $\beta=(V_{\rm circ}/V_{\rm hot})^{-\alpha_{\rm hot}}$, where 
$V_{\rm circ}$ is the circular velocity of the galaxy at the half-mass radius. The parameters $V_{\rm hot}$ and $\alpha_{\rm hot}$ are adjustable 
and control the strength of SNe feedback. The default model has $V_{\rm hot} = 485 \, {\rm km\,s}^{-1}$ and $\alpha_{\rm hot}$=3.2 \cite[cf.][]{Bower2006}. We removed the feedback strength of SNe by setting $V_{\rm hot}$=0 whilst keeping the photoionisation feedback.}. {Simply removing the feedback strength of SNe results in a model which greatly overpredicts the number of galaxies at all luminosities. 
In order to approximately reproduce the observations we modify the parameter in the Bow06 model which specifies the ratio between the sum of the mass in visible
stars and brown dwarfs, and the mass in visible stars. This parameter ($\Upsilon$) quantifies the assumption for the IMF of brown dwarfs (m $<$ 0.1M$_{\odot}$) which contribute mass but no light to stellar populations. We adopt a value of $\Upsilon$=4 for the NOSN and NOSN (no suppression) models.} More details on these models are provided in \cite{Kim2012a}. {Note that the generated Q$_{\rm cell}$ values used in the large volume simulations for each of the models in Table~\ref{Parameters} were calculated based on the Millennium-II dark matter simulation merger trees.} 

\begin{table*}
\caption{
The values of selected parameters which are different in the models. The columns are as follows: (1) the name of the model, (2) the value of the photoionization parameter $V_{\rm{cut}}$ (the suppression
of cooling occurs by the photoionisation feedback when the host halo's circular velocity lies below
a threshold value, $V_{\rm cut}$), (3) the SNe feedback parameter, $V_{\rm{hot}}$, 
(4) the IMF of brown dwarfs $\Upsilon$ (brown dwarfs contribute mass but no light to stellar population), and (5) comments giving model source or key differences from published models.}
\label{Parameters}
\begin{tabular}{lcccl}
\hline
\hline
 & $V_{\rm{cut}}$[kms$^{-1}]$&$V_{\rm{hot}}$[kms$^{-1}$]& $\Upsilon$ & Comments\\
\hline
Bow06 & 30  & 485  & 1 & Bower et~al. (2006), $V_{\rm{cut}}$ value change \\
Lagos11& 30  &  485  & 1   & Lagos et~al. (2012)\\
NOSN& 30 & 0 & 4       & Bower~et al. (2006), No SNe feedback \\
NOSN(no suppression)&0& 0 & 4 & Bower~et al. (2006)\\
 & & & & No SNe feedback and No photoionization feedback\\
\hline
\end{tabular}
\end{table*}

\begin{figure}
\begin{center}\vspace{-0.cm}
\includegraphics[width=8.6cm]{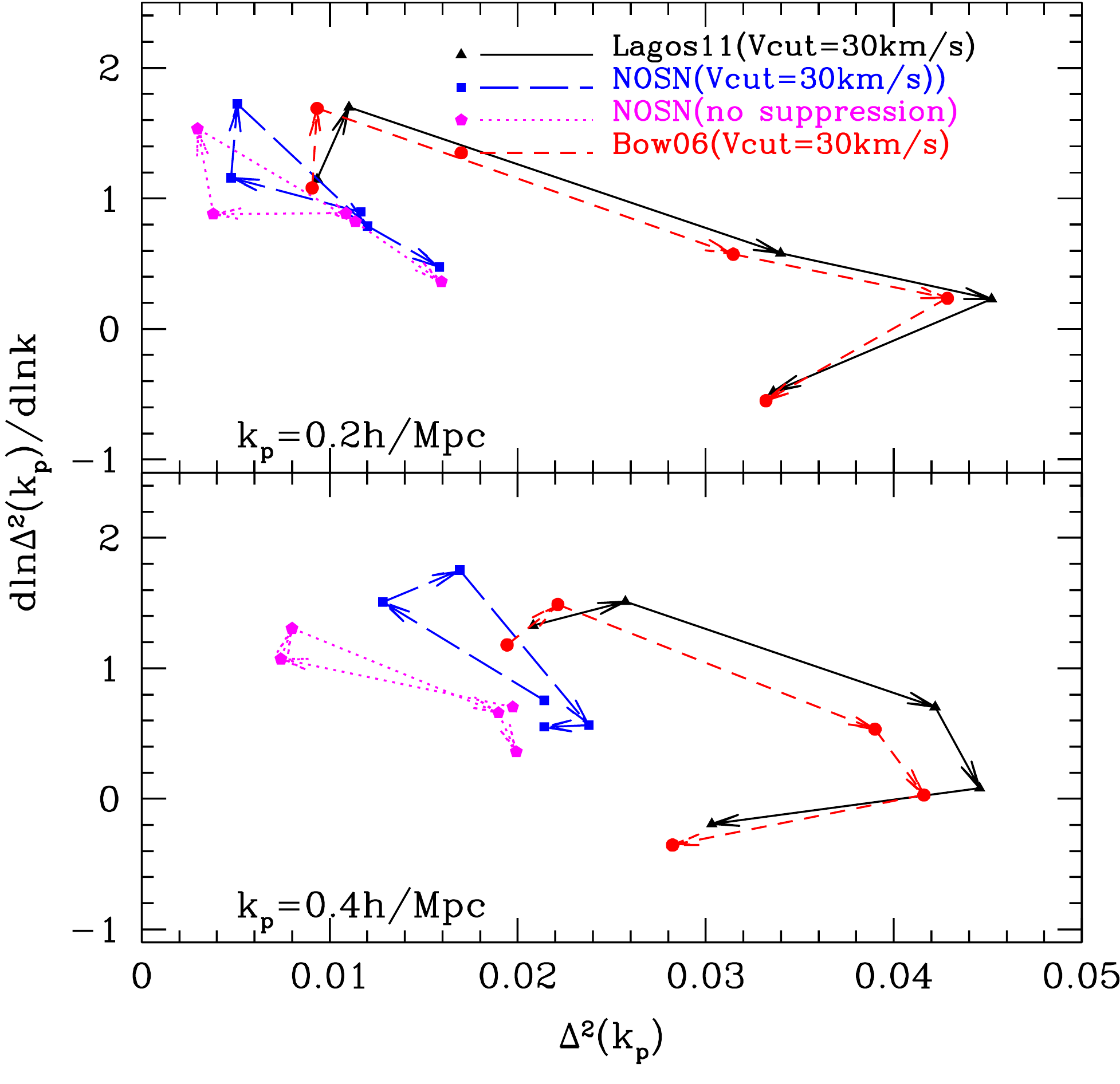}
\end{center}
\vspace{-0.2cm}
\caption{Plots show how the 21cm power spectrum changes using the loci of points in the parameter space of 21cm power spectrum amplitude and slope. Loci are shown for each of Lagos11 (our default model) (triangles, black solid line), NOSN(no suppression) (pentagons, violet dotted line), NOSN(V$_{\rm cut}$=30km/s) (squares, blue long dashed line), NOSN (no suppression) (octagons, green dot dashed line) and Bow06 (circles, red dashed line) models within the Millennium simulation. Results are shown for two central wavenumbers, $k$$_{p}$ =0.2$h$/Mpc (top) and 0.4$h$/Mpc (bottom), corresponding to the point on the power spectrum where we measure the amplitude and slope. Arrows show the direction from high to low {expected mean global mass averaged} neutral hydrogen fraction, $\left<x_{\rm HI}\right>$
}\label{DKLFMI}
\end{figure}

%\vspace{-1cm}
In each case we computed the 21cm power spectrum and plot the progression of a model in the parameter space of 21cm power spectrum
amplitude and slope ({note that since the ionized hydrogen fraction is not a direct observable}).
These curves are shown for the four models in Fig.~\ref{DKLFMI}, for wavenumbers $k$$_{p}$=0.2 (top) and 0.4$h$/Mpc (bottom). 
We also include arrows which show the direction from
high to low {expected mean global mass averaged} neutral hydrogen fraction, $\left<x_{\rm HI}\right>$ (from $\left<x_{\rm HI}\right>$=0.944 to 0.25; i.e., z=9.278 to 6.712). We see that the tracks separate into different parts of the plain, primarily according
to whether SN feedback is included or not (Bow06 and NOSN(V$_{\rm cut}$=30km/s)) 

{The regulation
of star formation and cooling of hot gas in small galaxies
by the SNe feedback process leads to massive galaxies which are
more biased towards dense regions, dominating the production of
ionizing photons. As a result, the amplitude of the redshifted 21cm power spectrum from the Bow06 model is larger than the NOSN (V$_{\rm cut}$=30km/s) model}. There are also small differences from the form of the star
formation law (Bow06 and Lagos11). {This is because the
modified star formation law in the Lagos11 model relative to the
Bow06 model leads to different predictions for the number of luminous
galaxies, and hence the clustering of the ionizing
source population.} {There are further small differences according to} whether photoionisation feedback is included or not [NOSN(V$_{\rm cut}$=30km/s) and NOSN(no suppression)]. {The NOSN(V$_{\rm cut}$=30km/s) model has a larger amplitude for the 21-cm power spectrum than does the NOSN(no suppression)
model because the photoionisation feedback effect in the absence of SNe feedback leads to more biased ionizing sources, so that the clustering amplitude increases.} Note that we do not include a model which has SNe feedback but no photoinoization feedback, because there is very little effect from photoionisation feedback in models which have SNe feedback \citep{Kim2012a}.
Fig.~\ref{DKLFMI} demonstrates that the power spectrum can be used to probe galaxy formation during reionization because the loci of the models fall in different parts of the parameter space of these observables. 

\section{Summary and conclusions}\label{Summary}
The ionization structure of the IGM during reionization, and hence the observed 21-cm power spectrum, will be sensitive to the astrophysical properties of the reionizing galaxies. Theoretical models which aim to describe reionization are challenged by the very large range of spatial scales involved. In particular, to understand and predict upcoming observations that come from the new generation of wide field telescopes, MWA, LOFAR, PAPER and SKA, large volume reionization simulations which cover an area comparable to or in excess of the field of view of telescope will be required. To address this problem, we extend the method described in \cite{Kim2012a} which connects galaxy formation and reionization using high resolution but relatively small volume N-body simulations. To calculate ionization structure in large volume simulations we use the relation between the distribution of ionisation fraction and dark matter overdensity to generate reionization maps within the Millennium, MXXL and GiggleZ-main simulations. 

We find that the amplitude of the redshifted 21-cm power spectra on large scales increases with simulation volume up to volumes of (500Mpc/$h$)$^{3}$ for $k$$<$0.1$h$/Mpc. The power spectra are converged at still larger scales. This implies that modelling within 0.5Gpc volumes will be sufficient for interpretation of forthcoming observes of the 21-cm power spectrum from reionization {$\sim$ $\left<x_{\rm i}\right>$=0.55. However since larger bubbles form in the highly ionized stage of reionization, we may need even larger volume simulations to see the convergence of the predicted 21cm power spectrum during the later stages of reionization $\left<x_{\rm i}\right>$$>$0.55}. 

We apply our simulations to explore the sensitivity of the 21cm power spectrum to the physics of galaxy formation. We find that measurements of the amplitude and slope of the 21-cm power spectrum will be able to determine the level at which SN feedback operated in high-redshift galaxies. Our method could be applied to any model of reionization {which has high resolution and sophisticated galaxy formation physics, but small volume, in order to interpret a large scale redshifted 21cm power spectra from upcoming observations}.

\vspace{5mm}

\hspace{-0.5cm}{\bf Acknowledgments} 
HSK is supported by a Discovery Early Career Researcher Awards from the Australian Research Council (DE140100940). The Centre for All-sky
Astrophysics is an Australian Research Council Centre of Excellence, funded by grant CE110001020. CMB acknowledges receipt of a Research Fellowship from the 
Leverhulme Trust. This work was supported in part by the Science and Technology Facilities
Council rolling grant to the ICC. The Millennium, Millennium II, and Millennium-MXXL Simulations were
carried out by the Virgo Consortium at the supercomputer centre of the
Max Planck Society in Garching.
Calculations for this paper were partly performed on the ICC Cosmology Machine, which is part of the DiRAC Facility jointly funded by STFC, the Large Facilities Capital Fund of BIS,
and Durham University.  

\newcommand{\noopsort}[1]{}

\bibliographystyle{mn2e}

%\bibliography{21cm-tosubmit}
\bibliography{21LARGE}

\label{lastpage}
\end{document}